\documentclass[preprint,review,12pt]{elsarticle}

\usepackage{float}
\usepackage{graphicx}
\usepackage{multirow}
\usepackage{color,amsmath,amssymb}

\newcommand{\2}{$_{2}$}
\newcommand{\cm}{cm$^{-1}$}
\newcommand{\etal}{{\textit{et al.}}}

\begin{document}

\title{A room temperature CO\2 line list with {\it ab initio} computed intensities}

\author{Emil Zak$^a$, Jonathan Tennyson$^a$\footnote{To whom correspondence should be addressed; email: j.tennyson@ucl.ac.uk}, 
Oleg L. Polyansky$^a$, Lorenzo Lodi$^a$}

\author{Nikolay F. Zobov$^b$}

\author{Sergey A. Tashkun$^c$, Valery I. Perevalov$^c$}

\address{$^a$Department of Physics and Astronomy, University College London,\\
London, WC1E 6BT, UK}
\address{$^b$Institute of Applied Physics, Russian Academy of Sciences, Ulyanov Street 46, Nizhny Novgorod 603950, Russia } 
\date{\today}
\address{$^c$V.E. Zuev Institute of Atmospheric Optics, SB RAS, 1, Academician Zuev Square, Tomsk 634021, Russia}
\begin{abstract}
  Atmospheric carbon dioxide concentrations are being closely
  monitored by remote sensing experiments which rely on knowing line
  intensities with an uncertainty of 0.5\% or better.  We report a
  theoretical study providing rotation-vibration line intensities
  substantially within the required accuracy based on the use of a
  highly accurate {\it ab initio} dipole moment surface (DMS).  The
  theoretical model developed is used to compute CO$_2$ intensities
  with uncertainty estimates informed by cross comparing line lists
  calculated using pairs of potential energy surfaces (PES) and DMS’s
  of similar high quality.  This yields lines sensitivities which are
  utilized in reliability analysis of our results.  The final outcome
  is compared to recent accurate measurements as well as the
  HITRAN2012 database.  Transition frequencies are obtained from
  effective Hamiltonian calculations to produce a comprehensive line
  list covering all $^{12}$C$^{16}$O$_2$ transitions below 8000
  cm$^{-1}$ and stronger than 10$^{-30}$ cm / molecule at $T=296$~K.

\end{abstract}

\maketitle

\newpage

\section{Introduction}

The quantity of carbon dioxide (CO$_2$) in Earth's atmosphere is thought to have a key role
in climate change and is therefore being closely monitored. Several agencies are flying experiments or whole
missions, for example
GOSAT \cite{GOSAT},  OCO-2 \cite{OCO} and ASCENDS \cite{ASCENDS},
to explicitly monitor the atmospheric CO$_2$ content.
Similarly, international ground-based networks such as TCCON \cite{TCCON} and NDACC \cite{NDACC} are
also dedicated to monitoring atmospheric CO$_2$ content.
A major aim of this activity is to establish CO$_2$ concentrations
at the parts per million (ppm) level
or, preferably, better. These projects will aim not only to look at
overall CO$_2$ concentration and its variation; it is of particular interest
to pinpoint where CO$_2$ is being
produced (sources) and where it is going (sinks).
This activity is clearly vital to monitoring and hopefully
controlling CO$_2$ and hence climate change \cite{Emmert2012}. 

All CO$_2$ remote sensing activities, both from the ground and space,
rely on monitoring CO$_2$ vibration-rotation spectra and therefore
are heavily dependent on laboratory spectroscopy for reliable
parameters; it is only through these parameters that atmospheric spectroscopic
measurements can be interpreted.  These spectroscopic parameters are of three
types: line centers, line profiles and line intensities. Line
centers or positions are established to high accuracy in many 
laboratory high
resolution spectroscopy studies and in general do not require significant
improvement for studies of Earth's atmosphere. Line profiles are more
difficult but significant progress on these has been made in recent
years with, for example, the inclusion of line mixing in both the
HITRAN database \cite{jt453} and many retrieval models, and the move 
beyond Voigt profiles \cite{jt584}. Here we focus on line intensities
for the main isotopologue of carbon dioxide, $^{12}$C$^{16}$O$_2$.

In the laboratory it is much harder to determine accurately line
intensities than line frequencies.  Typical accuracies for
experimental line intensity data used in atmospheric models and
retrievals is only 3 to 10~\%\
\cite{Wang2005,Perevalov2008,Song2010,Hashemi2013} and, until
recently, the best published measurements, e.g. Boudjaadar \etal,
\cite{06BoMaDa.CO2}, only provide accuracies in the 1 to 3~\%\ range,
still very significantly worse than the precision of 0.3 to 1~\%\
required by the modern remote sensing experiments
\cite{OBrien2002,XOCO,Sioris2014}.

Recently there have been a number of laboratory measurements aimed
at measuring absolute CO$_2$ line intensities with the high accuracy needed
for remote sensing \cite{07CaPaCa.CO2,09CaWeCa.CO2,11WuViJo.CO2,Hashemi2013,jt613,Malathy2015,15BrWeSe.CO2}.
With the exception of recent work by Devi \etal\ \cite{Malathy2015},
these studies have all focussed on obtaining the highest possible accuracy for a
few lines or even a single line. These investigations will be discussed further below. While
they
clearly do not provide the volume of data needed for remote sensing studies, they
do provide benchmarks that can be used to assess calculated intensities such
as those provided here. Approximately 20~000 transitions of $^{12}$C$^{16}$O$_2$ have been measured
experimentally; the experiments
up to 2008 were reviewed by Perevalov \etal\ \cite{Perevalov2008} and more recently
by Tashkun \etal\ \cite{Tashkun2015}.

There have been a number of attempts to use theory to provide
intensities for CO$_2$. Wattson \etal\
\cite{92WaRoxx.CO2,94DaMaBa.CO2} produced line lists using variational
nuclear motion calculations.  
More recently, Huang \etal\ have performed a series
of quantum mechanical studies giving line positions and intensities for CO$_2$
\cite{12HuScTa.CO2,13HuFrTa.CO2,14HuGaFr.CO2}. In particular, Huang \etal\
provide the most accurate currently available potential energy surface (PES)
for the CO$_2$ system.
A widely-used alternative theoretical
approach is based on effective operators for the
Hamiltonian and the spectroscopic dipole moment \cite{92TeSuPe.CO2}.
Currently, the effective Hamiltonian approach achieves one order of
magnitude better accuracy for $^{12}$C$^{16}$O$_2$ frequencies than
the best-available PES \cite{12HuScTa.CO2}. Within this framework, the
calculation of intensities requires eigenfunctions of an effective
Hamiltonian whose parameters were fitted to observed positions of
rotation-vibration lines as well as dipole moment operators tuned to
observed transition intensities.  This approach has been used to
create dedicated versions of the carbon dioxide spectroscopic databank
(CDSD) for room-temperature \cite{Tashkun2015} and high-temperature
\cite{02TaPeTe.CO2,11TaPe.CO2} applications.

Recently a number  of studies
have shown that it is possible to compute line intensities
using dipoles from {\it ab initio} electronic structure calculations with  an accuracy 
comparable to, or even better than, available measurements \cite{jt613,jt509,jt522,jt530,jt587}.
The intensity of a line depends on the transition line strength which
is obtained quantum-mechanically from the integral
\begin{equation}
 S_{if} = \left|\sum_\alpha \langle i | \mu_\alpha| f \rangle \right|^2
\label{linestrength}
\end{equation}
where here $|i\rangle$ and $|f\rangle$ are the initial and final state
rovibrational wavefunctions of the molecule and $\mu_\alpha$ is
component of the dipole moment surface (DMS).  The requirements for
accurate linestrengths are therefore high quality nuclear motion
wavefunctions and DMSs. Lodi and Tennyson \cite{jt522} developed a
procedure which provides estimated uncertainty on a
transition-by-transition basis based on the evaluation of multiple
line lists.  They initially applied this procedure to water vapor
spectra. Their data were used to replace all H$_2$$^{17}$O and
H$_2$$^{18}$O intensities for water in the 2012 release of HITRAN
\cite{jt557}. These data have since been critically assessed and
verified empirically for the 6450 to 9400 \cm\ region
\cite{14ReOuMiWa}. The present study combines the high accuracy {\it
  ab initio} DMS presented by Polyansky \etal\ \cite{jt613} and the
methodology of Lodi and Tennyson, which required some extension for
the CO$_2$ problem. This is discussed in the following section.

The current release of HITRAN \cite{jt557} takes its CO$_2$ line
intensities substantially from two sources: the Fourier transform
measurements of Toth \etal\ \cite{08ToBrMi} and an unpublished version
of CDSD \cite{12TaPe}.  The CDSD, whose intensities are
accurate to about 2 -- 20~\%\ depending on the vibrational band, has
recently been updated and released as CDSD-296 \cite{Tashkun2015}.
The uncertainty estimate 
is up to 20~\%\ for many transitions and is probably rather
conservative.
Recently some of
us computed a new, high accuracy DMS for CO$_2$ which we compared with
new high-accuracy experiments \cite{jt613} and the data in HITRAN. The
comparisons suggested that the new DMS is indeed excellent. In this
work we construct a new line list for $^{12}$C$^{16}$O$_2$ which we
suggest will significantly improve the precision of the intensity
parameters. Due to considerations associated with the DMS, this line
list is restricted to transition wavenumbers below 8000 \cm. However,
in this range the list should be comprehensive and includes
transitions which have yet to be quantified experimentally. The next
section presents the methodology used to construct the line list.
Section 3 presents the final line list and compares our results with
those from other sources. The final section gives our conclusions and
plans for future work.

\section{Methodology}

The Lodi-Tennyson method \cite{jt522} for validating linelists on a
purely theoretical basis relies on the use of accurate, {\it ab
initio} transition intensity calculations requires
an accurate procedures for obtaining nuclear motion wavefunctions
together with the use of at least two DMSs and two PESs. 
These aspects are described below.

\subsection{Ab initio surfaces}

The first stage in the molecular linelist evaluation process involves
computing energy levels and rotational-vibrational wavefunctions.
Our approach utilizes an exact nuclear kinetic energy operator
following the framework proposed by Tennyson and Sutcliffe
\cite{jt45,jt96,jt114,jt160} and implemented in DVR3D suite
\cite{jt338}; the quality of the electronic
PES provided is of primary importance. 
Energy levels and rotational-vibrational wavefunctions
obtained in this way are further used in intensity calculations,
requiring additionally a DMS function as  input. The accuracy of the
resulting line positions depends strongly on the quality of the PES, while
line intensities are dependent both on the PES and DMS. Therefore, in
order to generate high accuracy line intensities, it is necessary to
provide those two essential functions with the highest possible accuracy.
The present state-of-the-art {\it ab initio} PESs are capable of
reproducing experimental energy levels to 1 \cm\ accuracy, which still
remains insufficient for high resolution spectroscopy purposes.
Hence empirical fitting of {\it ab initio} surfaces has become a standard
procedure. This semi-empirical approach is much less successful in
the case of DMSs, partly due to technical difficulties in
obtaining accurate experimental data, suggesting the use of {\it ab initio}
DMSs is a better choice \cite{jt156}.  It is
natural to ask how different PESs and DMSs affect energy levels and
line intensities. Answering this, in turn, can shed some light on
the reliability of line intensities provided by our theoretical scheme.
Accordingly, the present study involves 6 independent runs of nuclear
motion calculations using the inputs presented below. \\

\textbf{Ames PES} 

As a primary choice we decided to use the 
semi-empirical Ames-1 PES from Huang \etal\ \cite{12HuScTa.CO2}, 
which is probably most accurate available. 
The fit of this PES started from a series of CCSD(T) {\it ab initio}
calculations with scaled averaged coupled-pair function (ACPF)
corrections, which also accounts for relativistic effects. No
non-Born-Oppenheimer effects were included, resulting in an 
isotope-independent PES. In addition to this a two-step refinement was
performed: first using a subset of HITRAN2008 $J=0-4$ energy levels,
second with the use of fully experimental levels for chosen $J$'s up
to 85.  The resultant PES was then rigorously tested against
HITRAN2008 and HITRAN2012  as well as against more recent experiments
\cite{12HuScTa.CO2,14HuGaFr.CO2}.  The best fit
root-mean-square-deviation (RMSD) with respect to purely experimental
energy levels for the final Ames-1 PES for the CO$_2$ main isotopologue
was equal to 0.0156 \cm\ in $J=0-117$ range. Comparison with HITRAN2012
database frequencies gave an average overall shift of $-0.0456$\cm\ and a
spread (rms) of 0.0712 \cm, which is consistent with our own calculation
based on this PES. The relatively large descrepancy between Ames-296 and
HITRAN2012 was the reason to exclude most of HITRAN energy levels from
the fitting procedure.  It also pointed to inconsistencies in the current
release of the database.

%  Even then, as further works revealed, 
% missing mass-dependent diagonal Born-Oppenheimer and non-adiabatic corrections,
% yielded an average RMSD with respect to HITRAN2012 database below 0.07 \cm
% \cite{14HuGaFr.CO2} for the isotopologue
%  band origins. This exceptionaly high 
% stability allows us to utilize single PES to all main isotopologues without
% significant loss of accuracy. 
% Indeed, 
% as previous experience shows, roughly 1 \cm PES error transfers into 0.5 \%
% error for strong and medium lines
%  \cite{jt509}, \cite{jt613}. Hence, by and large even 0.1 \cm PES isotopologue
% bias
% should influence well below our target accuracy in intensities. \\

\textbf{Ab initio PES} 

To aid the line sensitivity analysis, we independently constructed a fully {\it ab initio}
CO$_2$ PES using the energy points used by Polyansky \etal\  \cite{jt613}
to compute their DMS. MOLPRO 
\cite{12WeKnKn.methods} multi-reference configuration
interaction theory (MRCI) calculations with the aug-cc-pCVQZ basis 
were augmented by relativistic corrections at the
one-electron mass-velocity Darwin (MVD1)  level. 
For more details see the supplementary materials in ref. \cite{jt613}.
A fit with 50 constants to the MRCI grid points gave an 
RMSD of 1.54 \cm. The
relativistic correction surface was fitted separately with 31 
constants to yield a RMSD of  0.56 \cm.

A comparison with the Ames-1 PES shows a 1.5 \cm\ average discrepancy between the energy
levels computed with the
two surfaces for levels below 4000 \cm. Above this value some
energy levels spoil this
relatively good agreement to give a RMSD of 6.2 \cm\ for states below 11 000 \cm,
with 200 (0.5\%\ total) levels unmatched. However, for a fully {\it ab initio}
procedure this PES represents roughly the state-of-the-art for CO$_2$. It was therefore
used as part of the theoretical error estimation procedure.\\

\textbf{Fitted PES} 
Higher quality can be achieved by refining our {\it ab initio} PES with
Ames energy levels. This was done for levels with $J= 0, 1$ and 2. 
This fit resulted in a RMSD of 0.2 \cm\ between respective low $J$
energy levels and around 1.4 \cm\ RMSD for states including all $J$'s
(0-129) below 11 000 \cm, leaving only 30 levels above 10 000 \cm\
(0.1\%\ total) unmatched. 
% This PES was constructed to to make it closer
%to Ames-1 PES so to try to provide reliable estimate for line
%sensitivity. Large energetic deviations resulting from the use of the {\it
%ab initio} PES could result in artifical instability being shown by
%some lines, which in consequence would invalidate our error analysis. \\

\textbf{Ames DMS} 

The Ames dipole moment surface 'DMS-N2' was based on
2531 CCSD(T)/aug-cc-pVQZ dipole vectors \cite{13HuFrTa.CO2}. The linear least-squares fits were performed with 30 000 \cm\
 energy cutoff and  polynomial expansion up to 16-th order with 969 coefficients, which gave a RMDS of 
 $3.2\times10^{-6}$ a.u.
and $8.0\times10^{-6}$ a.u. for respective dipole vector components. Comparison with recent experiments  \cite{13HuFrTa.CO2} and 
CDSD data leads to the general conclusion that the Ames DMS, while reliable,
 still does not meet requiremets for remote sensing accuracy.

\textbf{UCL DMS} 

Our dipole moment surface was calculated using the finite field
method. Both positive and negative electric field vector directions
were considered for the $x$ (perpendicular to molecular long axis) and
$y$ (along molecular long axis) components of the dipole moment,
requiring 4 independent runs for each {\it ab initio} point. Finally
the dipole moment was computed as first derivative of electronic
energy with respect to a weak uniform external electric field
($3\times10^{-4}$ a.u.); a two-point numerical finite difference
approximation was used.  Previous research suggests that in general
derivative method yields more reliable dipole moments than those
obtained from simple expectation value evaluation \cite{jt475}.
Randomly distributed {\it ab initio} points were then fitted with a
polynomial in symmetry adapted bond-lengths and bond angle
coordinates. This resulted in an expansion up to fifth order. Points
above 15~000 \cm\ were rejected from the fit, leaving 1963 points for
the $x$ component fitted with 17 constants giving a RMSD of
$2.25\times 10^{-5}$ a.u.; and 1433 points for the $y$ component
fitted with 19 constants giving RMSD of $1.85\times10^{-5}$ a.u. .

\subsection{Nuclear motion calculations}

Nuclear-motion calculations were performed using the DVR3D suite
\cite{jt338}. Symmetrized Radau coordinates in bisector embedding were
applied to represent nuclear degrees of freedom.  Rovibrational
wavefunctions and energy levels were computed utilizing exact kinetic
energy operator (in Born-Oppenheimer approximation) with nuclear
masses for carbon (11.996709 Da) and oxygen (15.990525 Da).

As a first preliminary step in our procedure basis set parameters were optimized with respect to 
energy levels convergence using the Ames-1 PES. The final set of parameters for Morse-like basis functions \cite{jt338,jt14},
 describing stretching and bending motions, was considered as $r_e=2.95\; a_0$, $D_e=0.30\; E_h$ and $\alpha=0.0085\; E_h$.
 These values were chosen in a careful scan of parameter space with convergence speed as a criterion.
The contracted DVR basis set associated with Gauss-Legendre
quadrature points consisted of 30 radial and 120 angular functions, respectively. The appropriate choice of basis set parameters 
allowed us to 
reduce the size of the basis needed to converge energy levels, hence speeding up calculations. 
The same set of parameters was used for 
rovibrational energies evaluation with \textit{ab initio} and fitted PESs.
 
At room temperature the highest initial energy level that can be
populated enough to give a transition above the $10^{-30}$ cm/molecule
intensity threshold is roughly 6500 \cm\ and $J=130$.  Therefore we could
potentially be interested in upper energy levels up to 14 500 \cm\ to
cover the 0 -- 8000 \cm\ wavenumber region.  However, the current,
2012, version of HITRAN only considers upper states up to 11 500 \cm\
for wavenumbers below 8000 \cm.  As our target is to cover all HITRAN
transitions, we keep only energy levels below 11 500 \cm, so that the
Hamiltonian matrix in the first (vibrational) step of the calculation
(program DVR3DRJZ) could be truncated at 1000.  It guaranteed $J=0$
energy levels (band origins) below 10 000 \cm\ to be converged at the
$10^{-6}$ \cm\ level and energy levels around 12 000 \cm\ 
 at the $10^{-5}$ \cm\ level.

The ro-vibrational part of the computation (program ROTLEV3b) took advantage of symmetry adapted symmetric top basis set truncated 
at $600\times (J+1) $ for $J=0-50 $, $300 \times (J+1) $ for 
$J=51-86 $ and $100\times (J+1) $ for $J=87-129 $. 
This yielded 42~691 relevant\footnote{contributing to at least one transition with line intensity greater than $10^{-38}$ cm/molecule} energy levels up to 11 500 \cm\ and covered all HITRAN2012 database energy levels
 contributing to transitions up to 8000 \cm\ and $J\leq 129$.

The final step involved running the DIPOLE program \cite{jt338}. 
A uniform $10^{-30} $ cm/molecule cutoff value is sufficient to cover most of 
experimentally available data and also corresponds to HITRAN2012 standard, facilitating further comparisons. 
The value for the partition function  at 296~K $Q=286.096$
was taken from Huang et al.~\cite{14HuGaFr.CO2} and coincides with the value $286.095$  
obtained from the present calculation. 
For $^{12}$C$^{16}$O$_2$,  half of the possible energy levels do  not exist due to  nuclear spin statistics.
Transition intensities in cm/molecules were calculated using

\begin{equation}
 I(\omega)=4.162034\times10^{−19}~\omega_{if}g_iQ^{-1}(T)\left[\exp\left(\frac{E_i}{kT}\right)-\exp\left(\frac{E_f}{kT}\right)\right] S_{if}
\label{eq:Int}
\end{equation}
where $\omega_{if}$ is the transition frequency between the $i$'th and
$f$'th state, $g_i=(2J+1)$ is the total degeneracy factor, $Q(T)$ is the
partition function and $S_{if}$ represents the linestrength, see
eq.(\ref{linestrength}), for transition $i$ to $f$. Units for line
intensity are cm/molecule.

\subsection{Estimatation  of  the intensity uncertainties}

The dominant source of uncertainty in line intensities is given by the
{\it ab initio} DMS. The accuracy of the UCL DMS was considered in
detail by Polyansky \etal\ \cite{jt613} who suggested that for the
vast majority of transitions below 8000 \cm\ it should give
intensities accurate to better than 0.5~\%.

A characteristic of an {\it ab initio} DMSs is that entire
vibrational bands are reproduced with very similar accuracy.
This is because to a significant extent ro-vibrational transitions in a molecule
like CO$_2$ can be thought of as the product of
a vibrational band intensity and a H\"onl-London factor. Although DVR3D does not
explicitly use H\"onl-London factors, the use of
an exact nuclear motion kinetic energy operator ensures that these rotational
motion effects are accounted for exactly.

The nuclear motion wavefunctions give a secondary but, under certain
circumstances, important contribution to the uncertainties.
Variational nuclear motion programs yield very highly converged wavefunctions
and in situations where the PES is precise the 
intensities show little sensitivity to the details of how they are calculated.
For example, our wavefunctions calculated
using Radau coordinates give intensities very similar (to within 0.1~\%) to
those computed in the previous study \cite{jt613}
using Jacobi coordinates and different basis set parameters.

Where the wavefunctions do play an important role
is in capturing the interaction between different vibrational states. Such
resonance interactions can lead to intensity stealing
and, particularly for so-called dark states, huge changes in transition
intensities.
The Lodi-Tennyson methodology was designed to capture
accidental resonances which were not fully characterized by the underlying PES.
Under these circumstances calculations with
different procedures should give markedly different results. Lodi and Tennyson
monitored the effects of changes to the PES and
fits of the DMS. The procedure does not yield an uncertainty as such, it simply
establishes which transition intensities are correctly 
characterized by the calculation and hence have an uncertainty reflecting the
underlying DMS, and which are not, in which case the
predictions were deemed as unreliable and alternative sources of intensity
information was recommended.

In other words, trustworthy lines should be stable under minor PES/DMS
modifications.  One problem with this strategy is that if the
alternate PES (or DMS) differs too much from the best PES then large
intensity variations can be found which do not reflect problems with
the best calculation.  This issue already arose in a study on HDO
\cite{jtDown} where the {\it ab initio} and fitted surfaces showed
significant differences.  For CO$_2$ our {\it ab initio} PES is
relatively inacurate and hence far from the high quality Ames-1 fitted
PES; it was for this  reason we constructed a third PES by performing our
own, light-touch fit.

Here we therefore follow the Lodi-Tennyson strategy \cite{jt522} but
constructed and evaluated six linelists utilizing the three different
PESs and two different DMSs introduced above.  For this purpose
three sets of nuclear-motion wavefunctions were produced: the first
based on the Ames-1 semi-empirical PES, second based on the UCL {\it
  ab initio} PES and the third on our new fitted PES.  Those three
sets of wavefunctions were combined with the two {\it ab initio} DMSs
described above, to give line intensities.  Having six linelists, the
next step was to match line-by-line pairs of respective linelists:
(Ames PES \& Ames DMS, Ames PES \& UCL-DMS)=(AA,AU), (UCL-{\it ab initio} \&
Ames DMS, UCL-{\it ab initio} \& UCL-DMS)=(UA,UU), (fitted PES \& Ames DMS,
fitted PES \& UCL-DMS)=(FA,FU).  This stage was straightforward,
yielding almost 100\%\ match as the linelists being compared differ
only in DMS, which does not affect energy levels.  The second stage
involved matching the Ames-PES based with UCL-PESs based linelists, i.e.
(AA,AU) vs. (UA,UU) and (AA,AU) vs. (FA,FU). In both cases
line-by-line matching was preceeded by matching of energy levels.  In
the case of Ames vs. UCL we managed to match 90\%\ of lines stronger
than 10$^{-30}$ cm/molecule, while Ames vs. fitted resulted in high
99\%\ correspondence.  This confirms that reducing the 6.2 \cm\ RMSD
to 1.4 \cm\ makes a significant difference.  Note that since the (AU)
line list provides our best estimates of the intensities, there is no
benefit in performing a (UA,UU,FA,FU) scatter factor analysis.

\begin{figure}[H]
\begin{center}
\includegraphics[width=12cm]{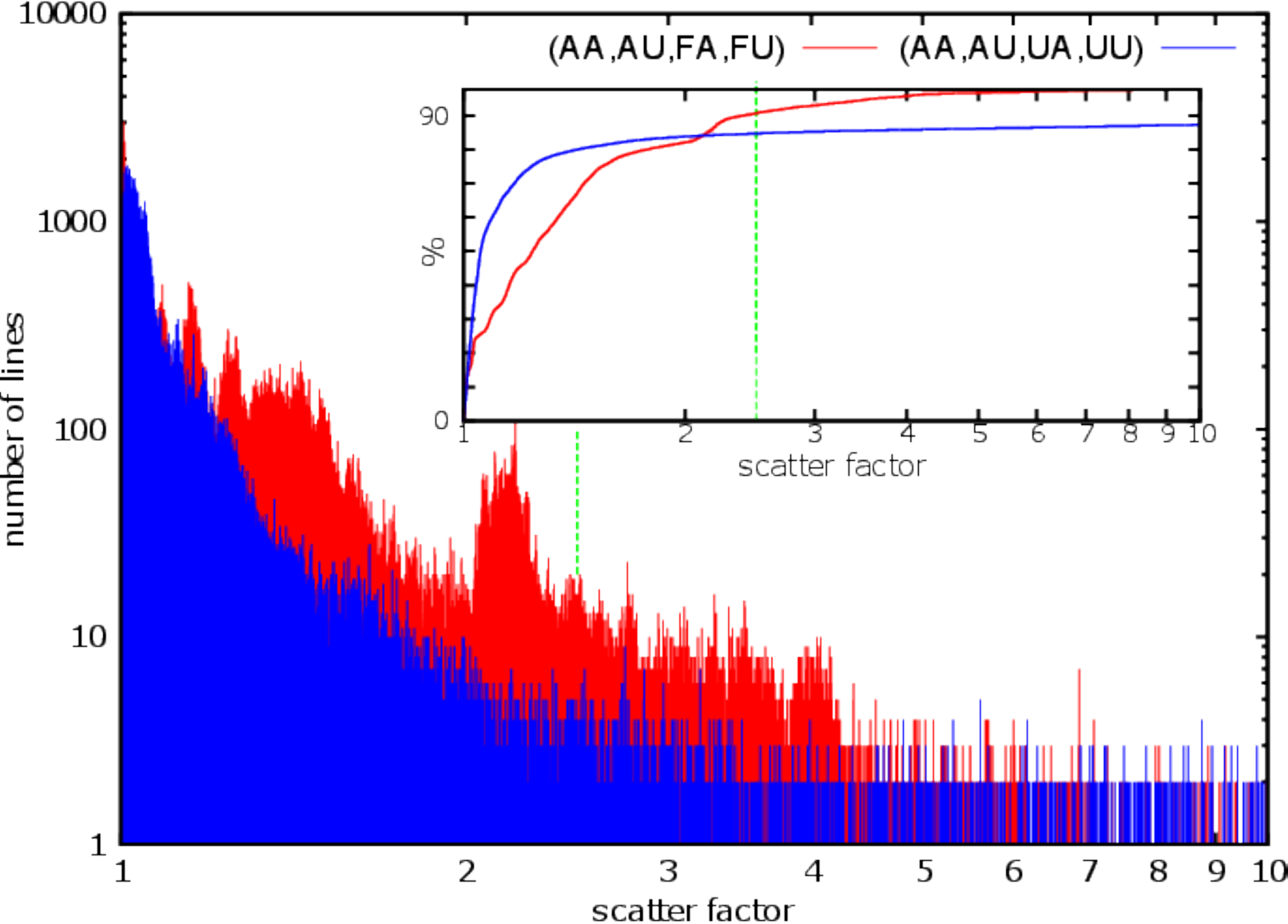}
\caption{Scatter factor, $\rho$, statistics for two sets of PES-DMS combination. Inset: cumulative distribution function. See text for further details.}
\label{figure:rho_stat}
\end{center}
\end{figure}

For each 'matched' line, the ratio of strongest to weakest transition intensity
was calculated, yielding a scatter factor $\rho$. 
Figure \ref{figure:rho_stat} shows scatter factors statistics for the two sets
of interest.
 We can clearly see that (AA,AU,UA,UU) set has more uniform and compact
distribution of $\rho$. 
However statistics for the {\it ab initio} UCL PES are based on an
incomplete match, leaving around 10\%\ of unmatched lines with an unknown
scatter factor. 
On the other hand, cumulative distribution function for (AA,AU,FA,FU) set reaches plateau at higher percentage of all lines,
indicating the advantage of fitted PES over UCL-{\it ab initio}. 

This leaves the 
problem of the choice of a critical value for the scatter factor. Lodi and
Tennyson chose the arbitrary value of $\rho = 2$.
Here we used the
scatter factor statistics to help inform our choice for this number. 
Figure \ref{figure:rho_stat} suggests $\rho=2.5$ 
is a reasonable value for this descriptor. Our more detailed analysis of
individual bands, given below, suggests that this is indeed
an appropriate value.

Detailed band-by-band comparisons revealed another feature of
(AA,AU,UA,UU) set: for a number of bands for which the AU intensities
gave excellent agreement with the measurements for all transitions,
but an arbitrary proportion of the transitions were identified as
being unstable. These false negatives are unhelpful and lead to the
risk of good results being rejected. For the (AA,AU,FA,FU) set we
found that provided the scatter factor was taken to be high enough,
$\rho > 2.5$, this problem was not encountered.  Hence we decided to
use fitted PES as a working set for further analysis.

For $J \geq 25$ it is quite common to have almost degenerate
transitions, that is transitions from exactly the same lower energy
level to upper energy levels with same $J$ and e/f symmetries, and as
close as 0.1 \cm. Therefore sometimes even the energetically best
match is not correct which leads to very inflated scatter factors.  In
this case, manual matching based on intensity considerations,
eliminates this problem for stronger bands ($I>10^{-26}$ cm/molecule)
and leaves only true $J$-localized instabilities.  Due to this issue with
almost degenerate transitions, we should note that the numerical
values of $\rho$ for unstable lines may in some cases be caused by
missasignments which leaked through our tests.  In particular, such a
situation can occur when  almost degenerate transitions 
have similar line intensities.

There are two main cases when \textit{ab initio} based intensities may
lose their reliability: energy levels crossing and intensity
borrowing by a weak band from a very strong band via resonance
interactions.

\begin{figure}[H]
 \begin{flushleft}
  \hfill\includegraphics[width=12cm]{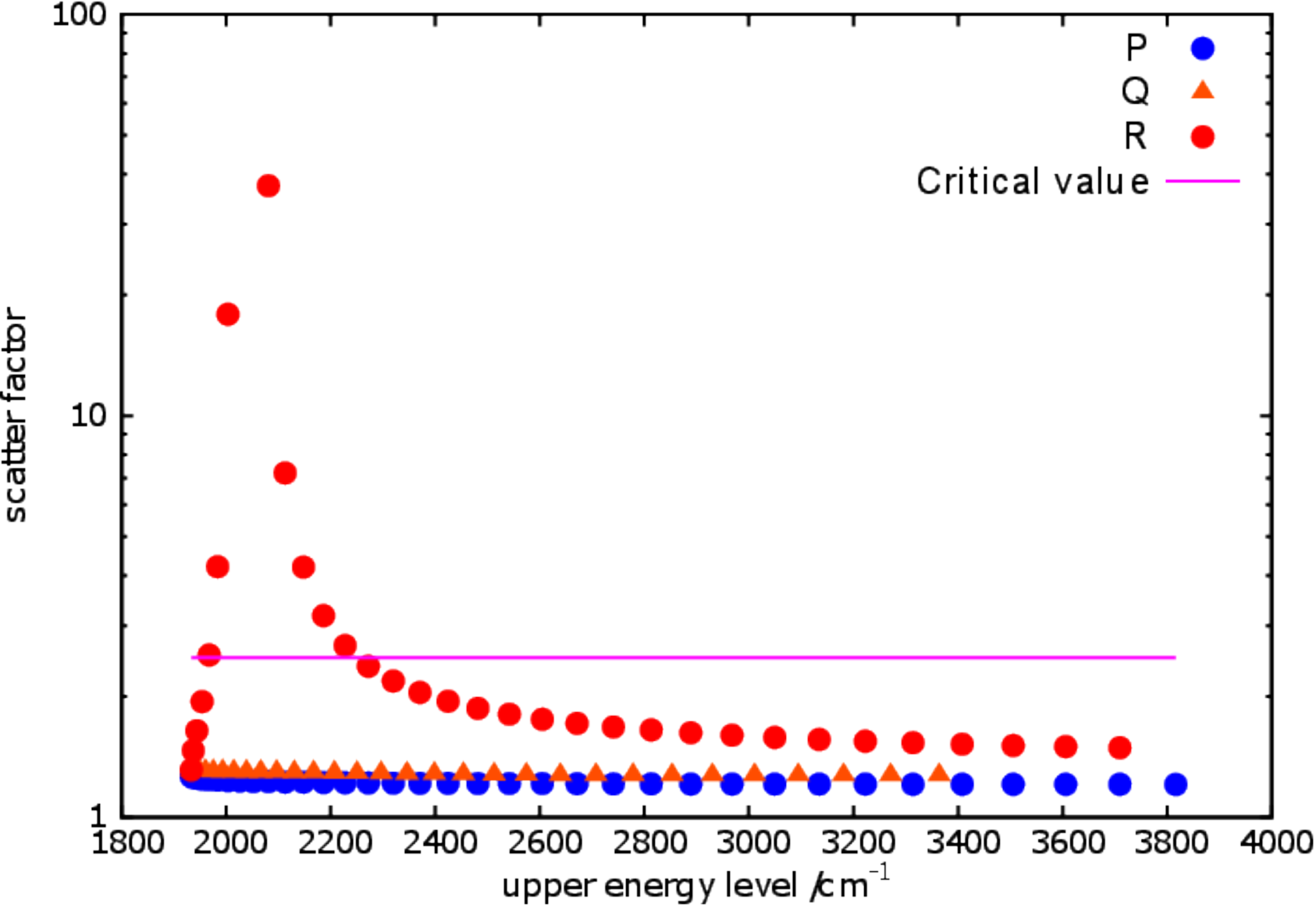}
%  \hfill\includegraphics[width=12cm]{ef_rho_11102-1.png}
\caption{Scatter factor as a function of lower energy level for the 
11102 -- 00001 band. The purple line denotes critical value of
the scatter factor ($\rho=2.5$). }
\centering
\label{figure:rho_11102}
 \end{flushleft}
\end{figure}

The latter is just the case for 1110$i$-00001 ($i=1,2,3$) bands.  They
borrow intensities from very strong asymetric stretching fundamental
via second order Coriolis interaction. This appears as a sharp peak
around 2000 \cm\ (upper energy level) as depicted in
Fig.~\ref{figure:rho_11102}. In this case, reproducing the line
intensities with high accuracy requires very precise wavefunctions. We
describe these lines as being associated with a $J$-localized
instability.

Together with bands for which scatter factor has peaks concentrated around certain energetic region
we also  encountered entire vibrational bands with
$\rho>2.5$, which we shall name as 'sensitive'.

The 40011-00001 and 40012-00001 bands are good
examples of combination of these two effects. Firstly both bands have
their upper energy levels around 8000 \cm\ and the whole bands are
uniformly unstable.  Moreover we observe peaking of the scatter factor
around $J= 76$ ($E_i$=2278 \cm), which we attribute to a strongly
mixed lower energy level involved in transition. The 40012-00001 band
is much stronger, therefore the $J=76$ transitions are still above our
intensity threshold, and we find a maximum in the scatter factor.

The Lodi-Tennyson approach was based on
the idea of occassional, accidental resonances. However it is well-known that
CO$_2$ has a series of systematic, Fermi-resonances.
We found that some of the bands gave large $\rho$
values for all transitions. There are two possible causes for this:
incomplete representation of the resonance interaction in the PESs used or
that the compared PESs differ significantly for this band.
Comparisons also suggested
that some of the predicted intensities for these bands may not be as reliable as one would
expect for the accurate UCL DMS. We therefore adjusted our strategy and introduced an intermediate
category of lines between stable and unstable for which the uncertainty of our intensity predictions
was increased.

\subsection{Line positions}

Lodi and Tennyson's water line list was based on the use of
experimental energy levels \cite{jt454,jt482,jt562} based on the
MARVEL procedure \cite{jt412,12FuCsi.method}.  For CO$_2$ an effective
Hamiltonian model was able to reproduce all published observed line
positions with accuracy compatible with measurement uncertainties
\cite{Tashkun2015}.  Specifically, the fitted model of $H_{\rm eff}$
was able to reproduce 44~917 observed line positions of
$^{12}$C$^{16}$O$_2$ having measurement uncertanties in the
$3.0\times10^{-9}$ \cite{04AmViCh.CO2}
to 0.02 \cm\ range with a dimesionless standard
deviation 2.0. This means that, on average, the obs-calc residuals
exceed the measured uncertainties by only a factor of two.  This makes
these calculatied line positions appropriate for our new
$^{12}$C$^{16}$O$_2$ line list.

\section{Results}

Our final line list given in the supplementary information,  includes the $\rho$ parameter, determined from
(AA,AU,FA,FU),
 as one of the fields; $\rho$ is set to $-1.0$  whenever  it could not be extracted. 
For the most intense bands this automatic procedure was followed by manual
matching and double-check, see Table~\ref{table:rho}.

\subsection{Scatter factors}

\begin{figure}[H]
  \includegraphics[width=10cm,angle=-90]{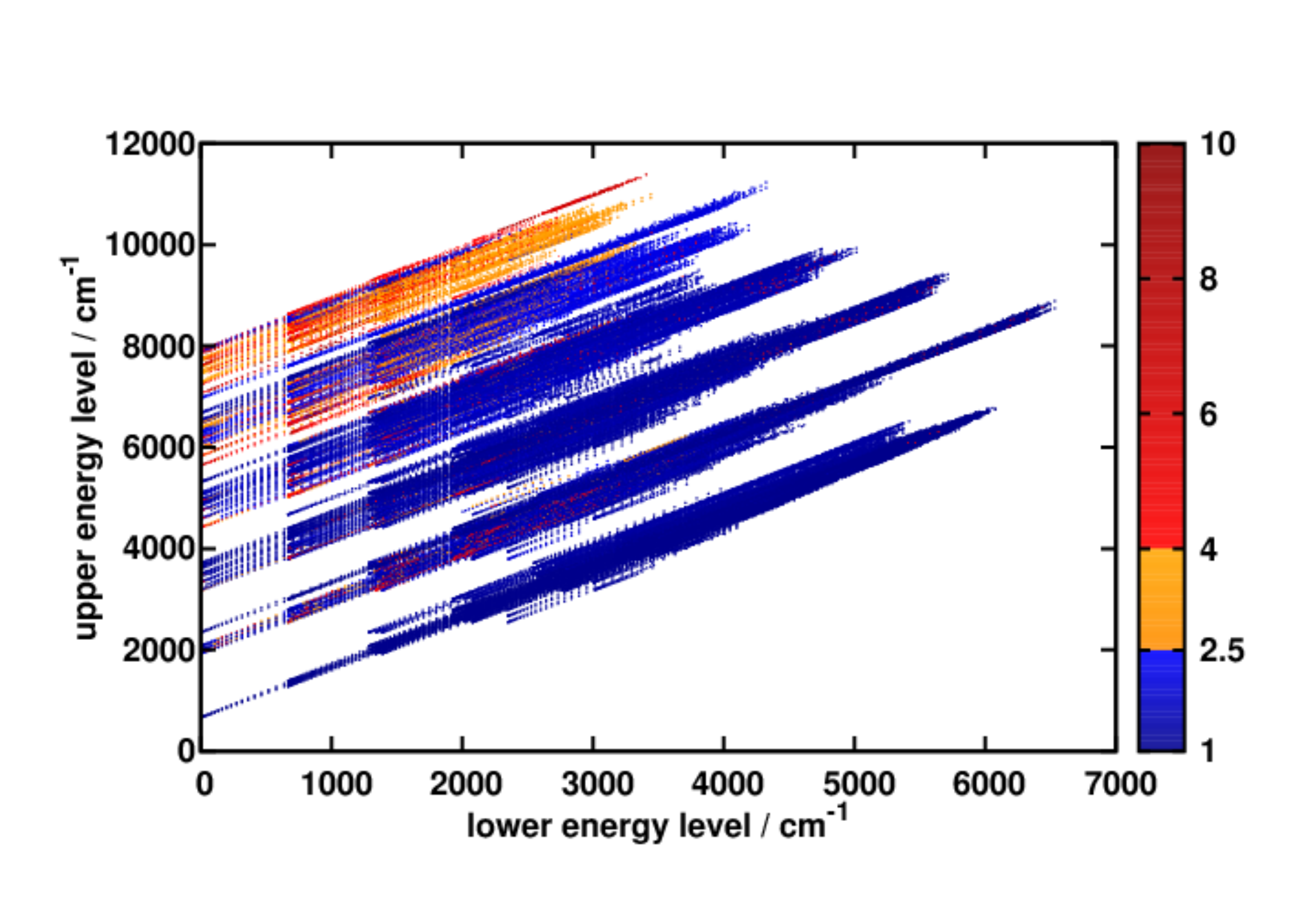}
\caption{Scatter factor map as a function of lower and upper energy level for transitions stronger than $10^{-30}$ cm/molecule. 
The color code represents the values of scatter factor, $\rho$. Four regions of 
line stability were determined:  blue-stable, orange-intermediate and red-unstable.  See text for further details}
\centering
\label{figure:rho_map}
\end{figure}

In order to appreciate the landscape of scatter factor distributions, it is
instructive 
to introduce scatter factor maps as a function of
lower and upper energy level. Figure \ref{figure:rho_map} shows a map where
color codes represent 
values of the scatter factor for a given transition. 
The fundamental bands are easily identified as straight lines originating at 0
\cm\ lower energy. 
The lowest hot bands originate at around 668 \cm, 
complicating the whole picture. A general conclusion from  figure 
\ref{figure:rho_map} is that the higher energy of a level involved in a
transition, the higher 
tendency for the transition to be unstable.  The color coding in the figure
divides 
scatter factor space into 3 regions of increasing instability, marked blue,
orange and red, 
respectively. The blue region is considered to be stable
and corresponding intensities are reliable. The orange region is intermediate
between stable and unstable, 
hence transitions marked orange need careful consideration.
The red region contains highly unstable lines whose computed line intensities
should not be trusted.  There are
a few super-unstable transitions ($\rho > 10$) which are not shown on the plots;
these lines are usually
associated with a strong 
resonance interaction with some other  energetically-close level. 
 Analysis of scatter factors for individual bands can yield insight.
By zooming in an energetic region of interest it is straightforward to pick up 
entirely unstable bands or single transitions which happen
to fall into resonance. Altogether we find 5400 transitions we classify
as unstable.

\begin{figure}[H]
  \includegraphics[width=14cm]{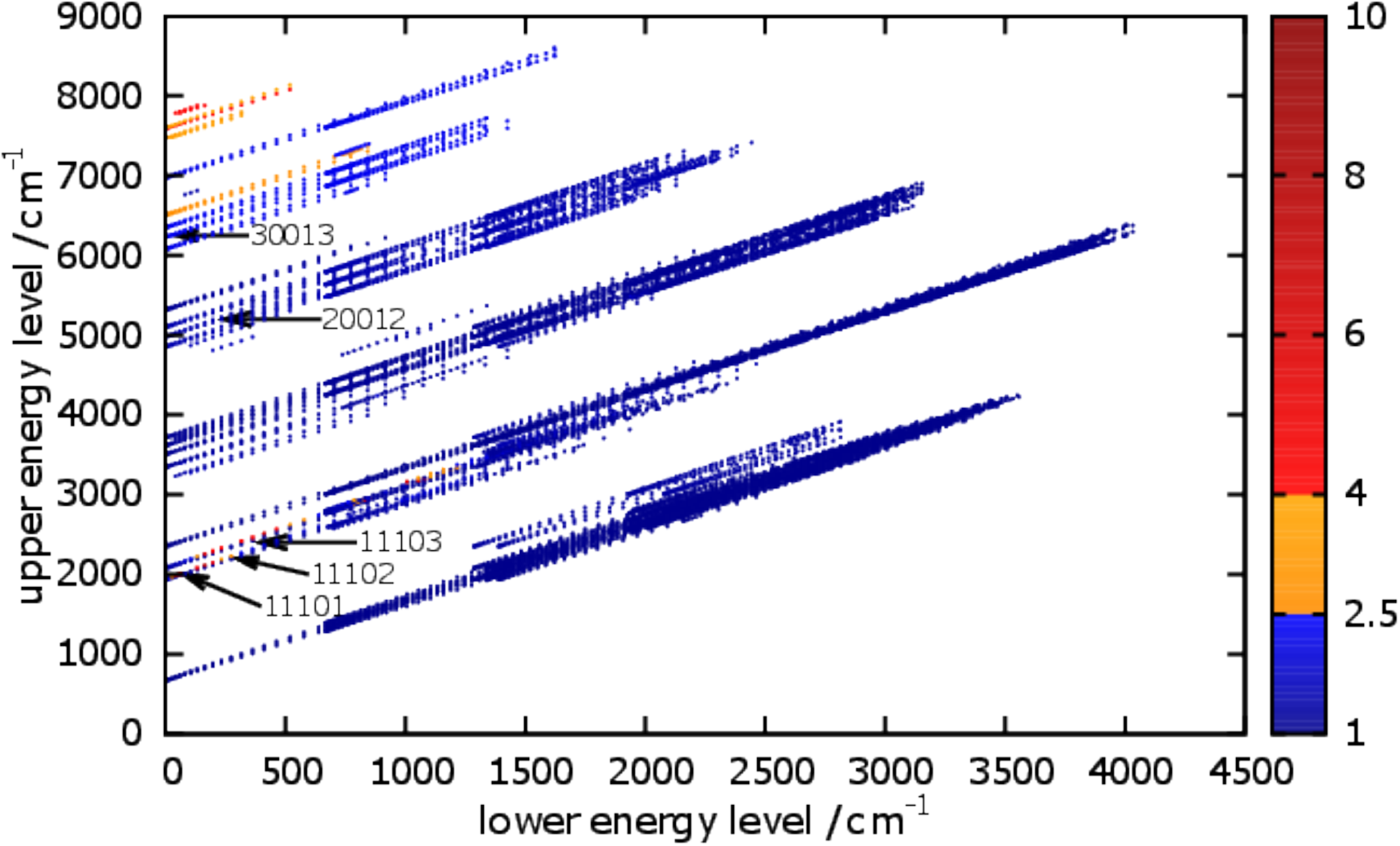}
\caption{Scatter factor map as a function of lower and upper energy level for transitions stronger than $10^{-25}$ cm/molecule. Color code represents the values of scatter factor. Four regions of 
line stability were determined: blue-stable, orange-intermediate and red-unstable. }
\centering
\label{figure:rho_map_25}
\end{figure}

For example, as can be seen from figure \ref{figure:rho_map_25} (which
considers only lines stronger than $10^{-25}$ cm/molecule) while
majority of bands have completely uniform scatter factors below the
critical value of 2.5, there are entire bands (marked orange)
systematically shifted by change of the underlying PES by a factor of
more than 2.5.  Those bands involving vibrational states which appear
to be influenced by strong resonance interactions are called
'sensitive' bands below. A completely different behaviour may be
observed for example for the 11101, 11102 and 11103 series of bands
(indicated with arrows). Here a fairly uniform scatter factor is
disturbed by $J$-localized peak.  Fig.~\ref{figure:rho_11102}
illustrates such behavior, which is explained by inter-band intensity
borrowing via rotational-vibrational (Coriolis) interaction terms in
molecular Hamiltonian.  A summary of stability analysis for selected
bands is given in table \ref{table:rho}.

\begin{table}[H]
\caption{Characterization of selected CO$_2$ bands. 
Given for each band are the band center in \cm, the total band strength in cm/molecule, the total number of lines in the band, the number of
stable lines with scatter factor $\rho < 2.5$, the number
of intermediate lines with $2.5 \geq \rho < 4.0$, the median of 
the scatter factor distribution
$\tilde{\rho}$, and the maximum and minimum value of $\rho$.}

\footnotesize
\begin{tabular}{l r l r r r r r r l}
\hline\hline
Band & Center & Strength & Total & Stable & Inter.&$\tilde{\rho}$ & $\rho_{max}$ & $\rho_{min}$& Type\\ [0.5ex] 
\hline 
%strongest
00011-00001 & 2349.949 & $9.20\times10^{-17}$ &129 & 129 & 0 & 1.0 & 1.0 & 1.0  & stable    \\ 
01101-00001 & 668.159 & $7.97\times10^{-18}$ &183 & 183 & 0 & 1.0 & 1.0 & 1.0  & stable    \\ 
01111-01101 & 2335.133 & $7.09\times10^{-18}$ &341 & 341 & 0 & 1.0 & 1.0 & 1.0  & stable    \\ 
10011-00001 & 3715.622 & $1.53\times10^{-18}$ &119 & 119 & 0 & 1.1 & 1.1 & 1.1  & stable    \\ 
10012-00001 & 3613.662 & $1.01\times10^{-18}$ &119 & 119 & 0 & 1.1 & 1.1 & 1.1  & stable    \\ 
02201-01101 & 669.309 & $6.15\times10^{-19}$ &340 & 340 & 0 & 1.0 & 1.0 & 1.0  & stable    \\ 
02211-02201 & 2321.865 & $2.71\times10^{-19}$ &317 & 317 & 0 & 1.0 & 1.0 & 1.0  & stable    \\ 
10012-10002 & 2328.264 & $1.73\times10^{-19}$ &115 & 115 & 0 & 1.0 & 1.0 & 1.0  & stable    \\ 
10001-01101 & 720.044 & $1.57\times10^{-19}$ &169 & 169 & 0 & 1.0 & 1.0 & 1.0 & stable   \\
10002-01101 & 617.239 & $1.46\times10^{-19}$ &169 & 169 & 0 & 1.0 & 1.0 & 1.0  & stable    \\ 
11111-01101 & 3721.742 & $1.21\times10^{-19}$ &310 & 310 & 0 & 1.1 & 1.1 & 1.0 & stable   \\
10011-10001 & 2327.419 & $1.04\times10^{-19}$ &113 & 113 & 0 & 1.0 & 1.0 & 1.0 & stable   \\ 
11112-01101 & 3578.816 & $7.58\times10^{-20}$ &309 & 309 & 0 & 1.1 & 2.2 & 1.0 & stable   \\
03301-02201 & 670.458 & $3.54\times10^{-20}$ &316 & 316 & 0 & 1.0 & 1.0 & 1.0  & stable    \\
20012-00001 & 4978.659 & $3.40\times10^{-20}$ &110 & 110& 0 & 1.4 & 1.5 & 1.3 & stable   \\ 
11102-10002 & 647.831 & $2.16\times10^{-20}$ &162 & 162 & 0 & 1.0 & 1.0 & 1.0  & stable    \\ 
11112-11102 & 2313.744 & $1.47\times10^{-20}$ &294 & 292 & 2 & 1.0 & 3.2 & 1.0  & stable, $J$-local     \\ 
11101-10001 & 689.438 & $1.36\times10^{-20}$ &159 & 159 & 0 & 1.0 & 1.0 & 1.0  & stable    \\ 
20011-00001 & 5100.494 & $1.10\times10^{-20}$ &107 & 107 & 0 & 1.4 & 1.5 & 1.3 & stable   \\ 
03311-03301 & 2308.597 & $1.03\times10^{-20}$ &291 & 291 & 0 & 1.0 & 1.0 & 1.0  & stable    \\ 
11111-11101 & 2312.260 & $7.23\times10^{-21}$ &290 & 290 & 0 & 1.0 & 1.0 & 1.0 & stable   \\
20013-00001 & 4854.447 & $7.13\times10^{-21}$ &109 & 109 & 0 & 1.5 & 1.5 & 1.5  & stable  \\   
11101-02201 & 740.173 & $6.14\times10^{-21}$ &308 & 308 & 0 & 1.0 & 1.0 & 1.0  & stable    \\ 
11102-02201 & 595.761 & $5.33\times10^{-21}$ &304 & 304 & 0 & 1.0 & 1.0 & 1.0  & stable    \\ 
11101-00001 & 2077.641 & $5.17\times10^{-21}$ &107 & 97& 3 & 1.9 & 1500 & 1.4  & stable, $J$-local    \\ 
12212-02201 & 3724.349 & $4.75\times10^{-21}$ &284 & 284 & 0 & 1.1 & 1.1 & 1.1  & stable    \\
20012-10002 & 3693.261 & $3.69\times10^{-21}$ &104 & 104 & 0 & 1.1 & 1.1 & 1.0  & stable    \\ 
20013-10002 & 3569.048 & $3.12\times10^{-21}$ &104 & 104 & 0 & 1.1 & 1.1 & 1.0  & stable    \\ 
20011-10001 & 3712.291 & $2.96\times10^{-21}$ &102 & 102 & 0 & 1.1 & 1.1 & 1.0  & stable    \\ 
04401-03301 & 671.607 & $1.80\times10^{-21}$ &290 & 290 & 0 & 1.0 & 1.0 & 1.0  & stable    \\ 
12202-11102 & 654.112 & $1.57\times10^{-21}$ &294 & 294 & 0 & 1.0 & 1.0 & 1.0  & stable    \\ 
00031-00001 & 6973.378 & $1.38\times10^{-21}$ &101 & 101 & 0 & 2.1 & 2.2 & 2.0  & stable    \\ 
00011-10001 & 961.746 & $9.01\times10^{-22}$ &99 & 99 & 0 & 1.2 & 1.2 & 1.2  & stable    \\ 
12201-11101 & 685.423 & $8.03\times10^{-22}$ &291 & 291 & 0 & 1.0 & 1.0 & 1.0  & stable    \\ 
11102-00001 & 1933.229 & $6.19\times10^{-22}$ &156 & 146& 3 & 1.4 & 37 & 1.2  & stable, $J$-local    \\ 
30011-00001 & 6503.913 & $5.17\times10^{-23}$ &24 & 0 & 24 & 2.6 & 2.6 & 2.6  & sensitive   \\
12201-01101 & 2094.904 & $5.01\times10^{-22}$ & 300 & 271 & 7 & 1.3 & 1200 & 1.1  & stable, $J$-local   \\ 
30013-00001 & 6228.740 & $4.54\times10^{-22}$ &99 & 99& 0 & 2.3 & 2.3 & 2.3  & stable   \\ 
30012-00001 & 6348.693 & $4.54\times10^{-22}$ &99 & 99 & 0 & 2.2 & 2.3 & 2.1  & stable    \\ 
20001-11101 & 719.501 & $3.89\times10^{-22}$ &146 & 146 & 0 & 1.0 & 1.0 & 1.0  & stable    \\ 
13311-13302 & 2490.039 & $9.13\times10^{-24}$ &75 & 10 & 65 & 2.5 & 3.5 & 2.4  & sensitive    \\
40012-00001 & 7735.305 & $3.19\times10^{-24}$ &24 & 0 & 24 & 2.6 & 2.6 & 2.6  & sensitive \\
40011-00001 & 7921.693 & $2.10\times10^{-25}$ &24 & 0 & 24 & 2.6 & 2.6 & 2.6  & sensitive \\ 
23302-22201 & 481.776 & $9.92\times10^{-26}$ &90 & 90& 0 & 1.0 & 1.0 & 1.0  & stable     \\
30004-11102 & 1859.407 & $6.77\times10^{-26}$ &24 & 0 & 24 & 2.6 & 2.6 & 2.6  & stable, $J$-local \\

\hline\hline
\end{tabular}
\label{table:rho}
\end{table}

108 out of 116 bands stronger than $10^{-25}$ cm/molecule are stable.
Bands involving bending excitations are also very stable.  For some
bands, such as 32203--03301 and 42201--03301 $J$-localized instabilities
appear only weakly, generating peaks which do not exceed the critical
value.

\subsection{Comparison with high-accuracy measurements}

Polyansky \etal\ \cite{jt613} showed that transition intensities
based on the (A,U) model gave excellent agreement with new, high
accuracy measurements of the 30013 -- 00001 band in the
6200--6258 \cm\ reported in the same paper. Polyansky \etal\ also
compared their predictions with the high accuracy mearurements
of Casa \etal\ \cite{07CaPaCa.CO2,09CaWeCa.CO2} and Wuebbeler
\etal\ \cite{11WuViJo.CO2} for the 20012 -- 00001 band.
While their results were in excellent agreement with the single line
intensity measured by  Wuebbeler
\etal, they suggested that the results of Casa \etal's results were
significantly less accurate than claimed. This assertion has since
been confirmed by new high-accuracy
 measurements performed by Brunzendorf \etal\ \cite{15BrWeSe.CO2} which
show almost no systematic shift and average deviation
 of only 0.35\%\ with respect to our 
(and Polyansky \etal's) predictions.

Two more lines in $\nu_1+\nu_3$ band (P34,P36) were measured by Pogany
\etal\ \cite{13PoOtWe.CO2} with reported 1.1 and 1.3 \%\ uncertainty.
The corresponding UCL intensities deviate by 2.0\% and 2.5\%
respectively. Nevertheless these are on average 1\% closer to
experimental values than the intensities obtained from either Ames-296
or CDSD-296.

% Recent measurements by Casa et al. reveal some discrepancies with respect to our calculations. As we shown in PRL paper....discuss. 
% I have numeric data from S. Tashkun - make some statistics on recent measurements.

\begin{figure}[H]
  \includegraphics[width=12cm]{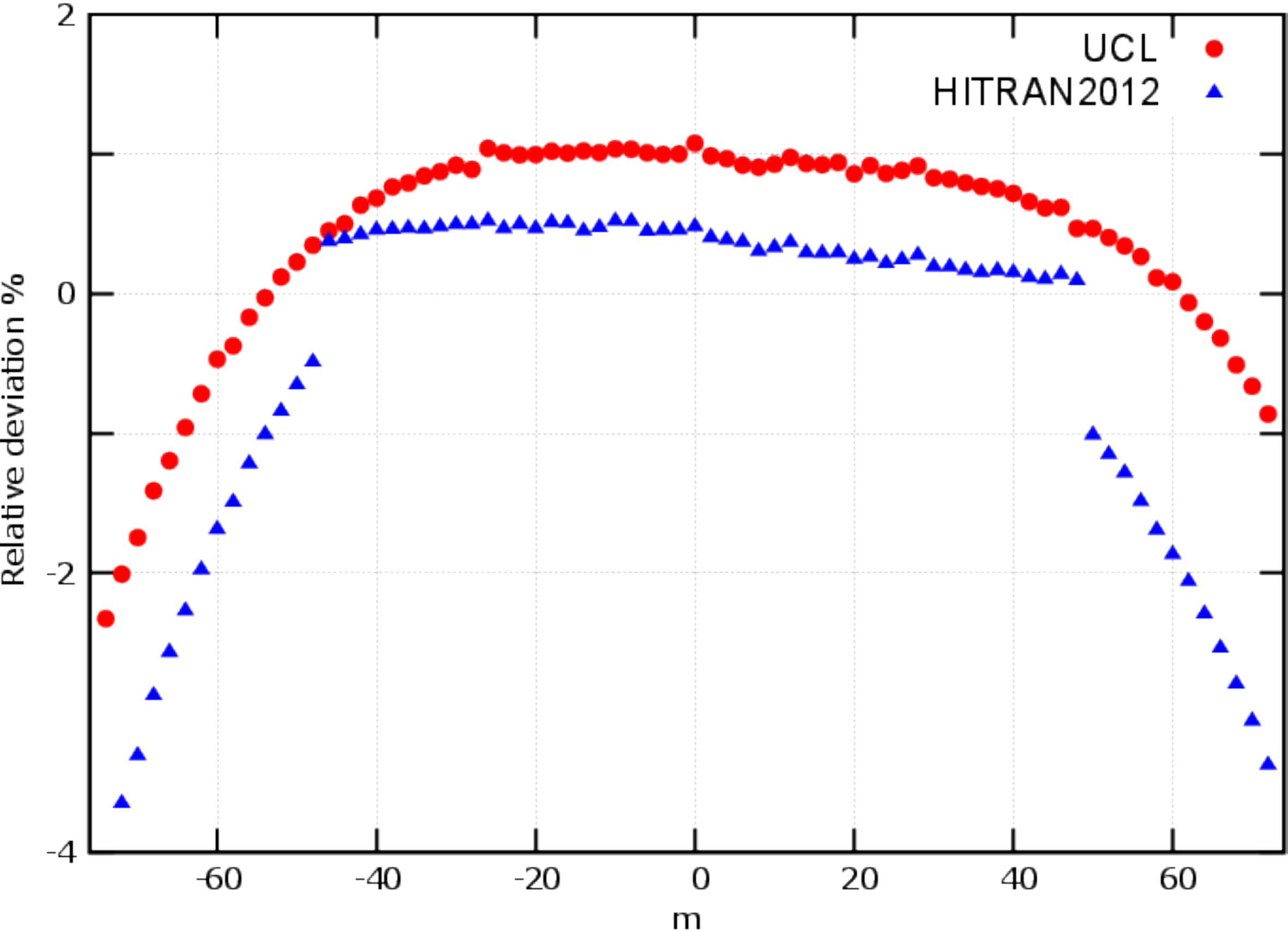}
\caption{Comparison of experimental line intensities from Devi \etal\ \cite{Malathy2015} for the 30013 -- 00001 band with present (UCL) and HITRAN2012 values. Relative deviation is defined as 
$\left[\frac{I(x)}{I(Devi)}-1\right]\times 100\% $, where $x=$HITRAN,UCL.}
\centering
\label{figure:Devi}
\end{figure}

Very recently Devi \etal\ \cite{Malathy2015} performed a detailed
study at 1.6 $\mu$m. The strongest band in this region is
30013 -- 00001. A comparison between their measured line intensities and our predictions is
given in figure \ref{figure:Devi}.

From figure \ref{figure:Devi}
 it is evident that a majority of our line
intensities lie within 1 \%\ of the new measurements. The bow-like
behavior seen particularly  at high $J$'s here, and in other
comparisons discussed below, is unlikely to be due
to our calculations. Instead we expect it is an
artifact associated with Herman-Wallis factors used to parameterize
the experimental data, which tend to
overestimate
line intensities for high-$J$ transitions. If this is so, then it is likely
that our results match Devi \etal's at the sub-percent level.

Low $J$ HITRAN2012 line intensities for the 30013 -- 00001 band
originate from the JPL OCO linelist of Toth \etal\ \cite{08ToBrMi} and
lie on average 0.5~\%\ above the value of Devi \etal. These lines are
marked with a 7 as the HITRAN intensity uncertainty code which means
that these line lintensities are accurate within 2\%. The high-$J$
line intensities ($J > 45$) are all calculated values based on a fitted
effective dipole moment model \cite{12TaPe}. They have 3 as
the HITRAN intensity uncertainty code but may have errors in the
intensity greater than 20\%. The clearly visible jump in HITRAN points
in figure \ref{figure:Devi} is located at the meeting point of the two
data sources.

\subsection{Comparison with other line lists}

\textbf{Ames-296}

Huang \etal\ \cite{14HuGaFr.CO2} published infrared line lists for 12
stable and 1 radioactive isotopologues of CO$_2$.  These linelists
were calculated with Ames-1 PES \cite{12HuScTa.CO2} and DMS-N2
\cite{13HuFrTa.CO2}, or (A,A) in our notation above.  
 We generated from their data a
$^{12}$C$^{16}$O$_2$ line list for its natural abundance, 
$T =296$~K and with an intensity cutoff of $10^{-30}$ cm/molecule, which
we refer to as Ames-296.  Ames-296 contains 162~558 lines
in the 0 - 8000 \cm\ range.  To facilitate comparison with other
line lists we performed a spectroscopic assignment of this line list.
As a first step, for the sake of consistency, it was necessary to
compare energy levels from original Ames-296 linelist with our DVR3D
recalculation.  Energy levels up to 6000 \cm\ gave a RMSD
of 0.05 \cm\, and 0.06 \cm\ up to 10 000 \cm. This is slightly more than
we would have expected  on the basis of previous comparisons \cite{jt309}
and appears to be due a slightly non-optimal choice integration grids
in Huang \etal's calculations (Huang and Lee, 2015, private communication).\\

\textbf{CDSD-296}

The effective operator approach enables one to reproduce all published
observed positions and intensities with accuracies compatible with
measurement uncertainties. Based on fitted $H_{\rm eff}$ and $D_{\rm
  eff}$ models Tashkun et al. \cite{Tashkun2015} created a high
resolution spectroscopic databank CDSD-296 aimed at atmospheric
applications.  The databank contains the calculated line parameters
(positions, intensities, air-and self-broadened half-widths,
coefficients of temperature dependence of air-broadened half-widths
and air pressure-induced lineshifts) of the twelve stable isotopic
species of CO$_2$. The reference temperature is 296 K and the
intensity cutoff is $10^{-30}$
cm/molecule. \\

\begin{figure}[H]
  \includegraphics[width=12cm]{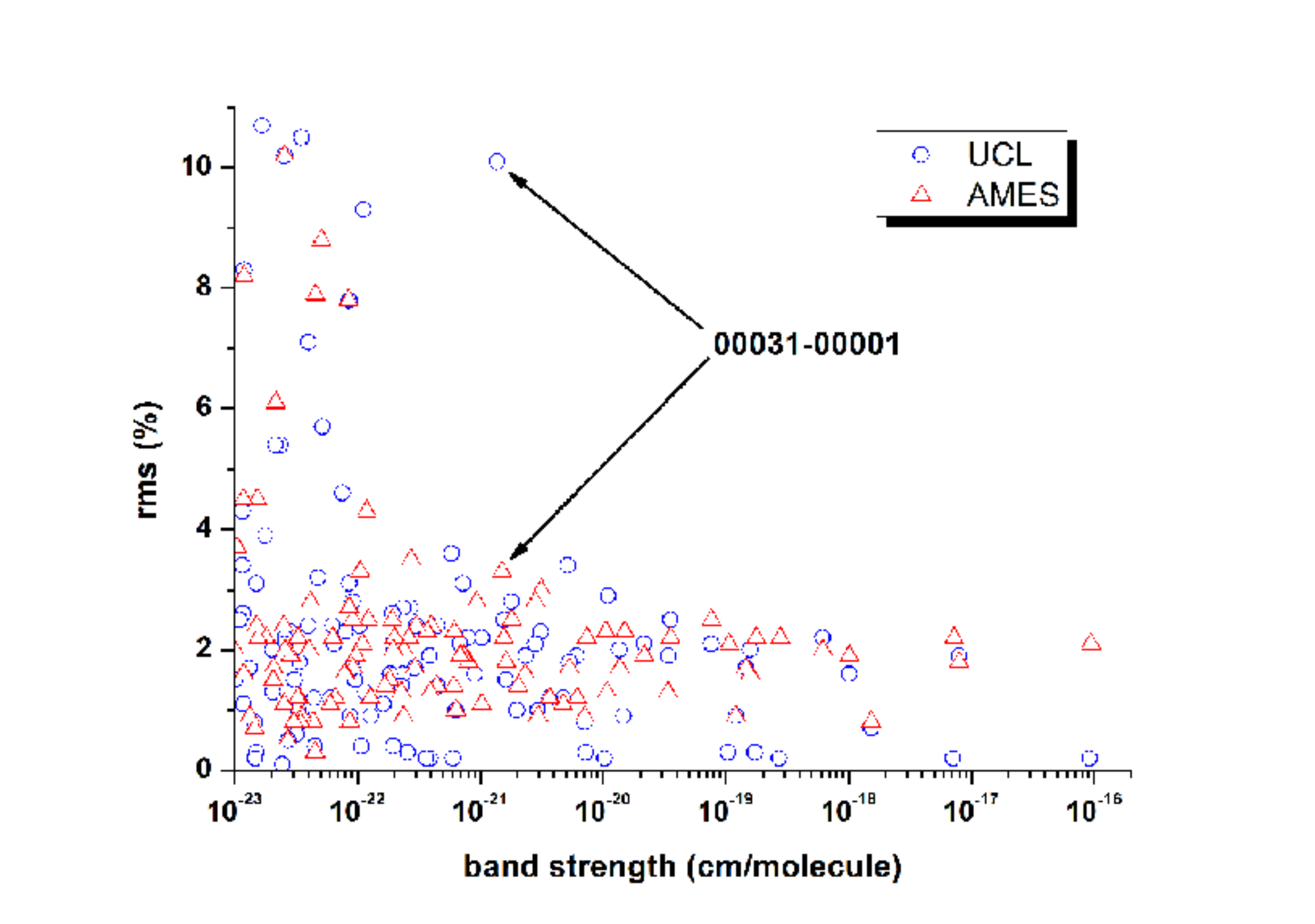}
\caption{Root mean square deviation for bands intensities of Ames-296 (red
triangles) and the present results (UCL, blue circles) with respect to CDSD-296}
\centering
\label{figure:CDSDrms}
\end{figure}

%  However, this is a difficult task since measurements were reported with different temperatures, abundances and units. 
Figure \ref{figure:CDSDrms} compares Ames and UCL line intensities with the
semi-empirical CDSD-296 results.  For the sake of clarity only strong
bands with intensities greater than $10^{-23}$ cm/molecule are
plotted.

For the strongest bands UCL linelist agrees much more closely with
CDSD-296 than Ames-296 does.  The only real exception to this are the
00031-00001 and 01131-01101 bands. For this family of bands, whose
intensity derives from the same dipole moment derivative, the
deviations from Ames-296 are three times less than UCL ones.
We identified 3170 transitions belonging to this family.

\subsection{HITRAN2012}

HITRAN2012 \cite{jt557} contains 160~292 $^{12}$C$^{16}$O$_2$
lines in 0 -- 8000 \cm\ region.  
A matching procedure for our Ames-1 PES
energy levels to those originally extracted from HITRAN2012 database
was conducted by imposing rigorous restrictions on rotational quantum
numbers and rotationless parities as well as 0.3 \cm\ tolerance for
energy difference. This scheme matched all 16~777
unique energy levels present in HITRAN2012 covering $J$ values from 0 to
129  with RMSD of 0.07 \cm.
The largest deviation found between two levels was roughly 0.2 \cm. 

The next step was to match transition lines between HITRAN2012 and UCL
linelists. The procedure relied on a simple algorithm, where
corresponding lines were matched using already matched energy
levels list. As a result all 160~292 lines up to 8000 \cm\ were matched
with a RMSD of 0.08 \cm\ in line positions.

There are two main sources of HITRAN2012 data for CO$_2$ main
isotopologue: a small set of 605 lines in 4800-6989 \cm\ range
originating from experiment (JPL OCO line list) by Toth \etal\
\cite{08ToBrMi} and the majority of transitions from a previous
version of CDSD.  In general data from latest version
of CDSD-296 are very close to line positions and intensities given in
HITRAN2012.

\begin{figure}[H]
\hfill\includegraphics[width=12cm]{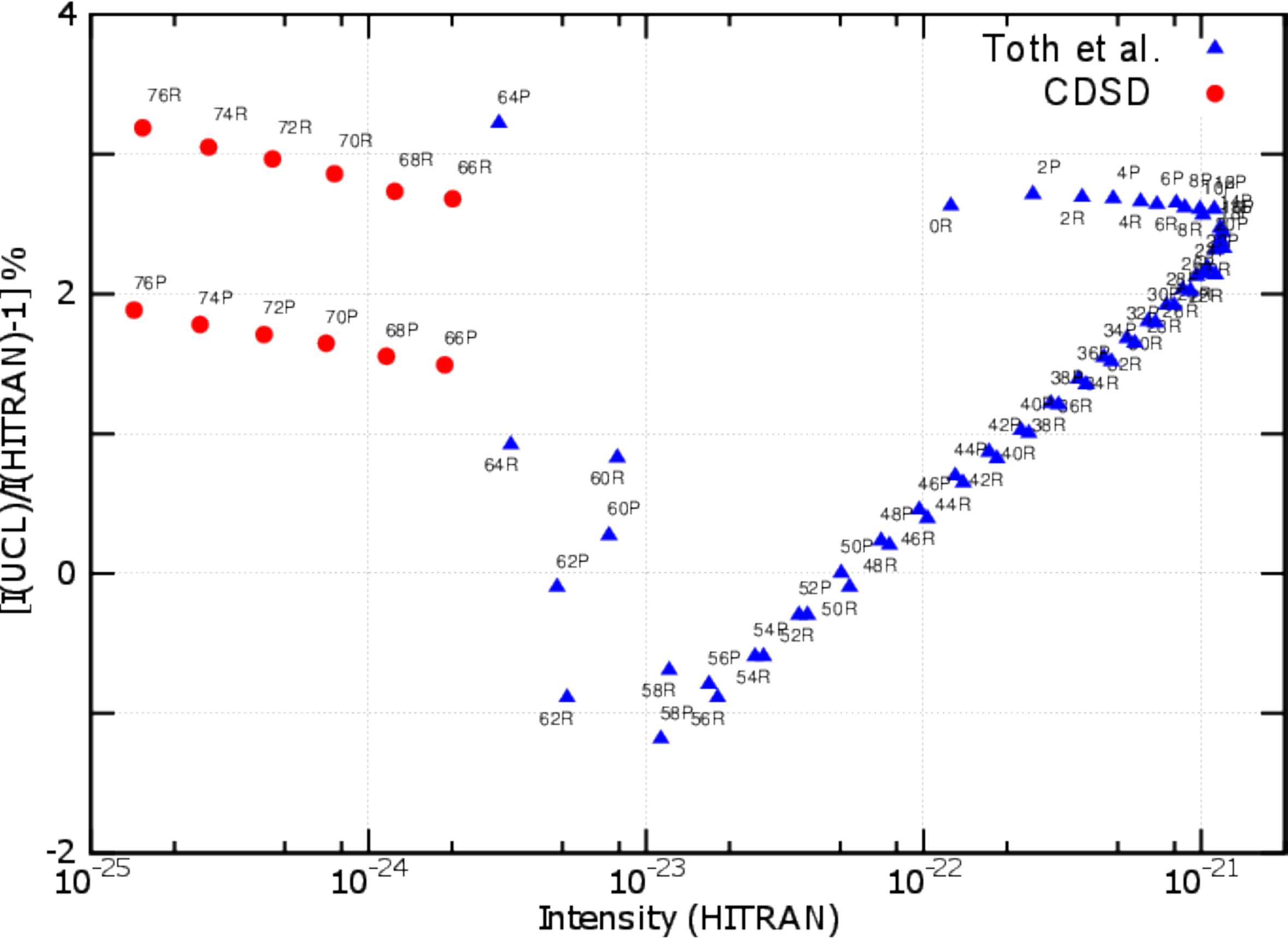}
%\hfill\includegraphics[width=12cm]{Int_20012-1.png}
\caption{HITRAN2012 vs. UCL line intensities comparison for the 20012 -- 00001 band. Two HITRAN data sources are marked with red circles (CDSD-semiempirical calculations) and blue (Toth \etal-experimental)}
\centering
\label{figure:Int_20012}
\end{figure}

The estimated uncertainties for all CDSD intensities is given as 20
\%\ or worse in HITRAN (uncertainty code 3). On the other hand, Toth \etal's
intensities are supposed be accurate to better than 2\%\ (uncertainty
code 7) or 5\%\ (code 6). This reveals two
issues with current version of HITRAN: \\
a) The stated uncertianty estimate
of all current entries are
insufficiently 
accurate for remote sensing applications. Our previous study \cite{jt613} already
showed that for a number of important bands the actual accuracy of the
intensities in HITRAN is much higher than suggested by their estimated uncertainties.\\
 b) line intensity accuracies
are not uniform throughout the spectral region. 
Our experience from studies on several molecules is that the ratio of observed
to variational line intensies should be roughly constant for a given unless there
is an isolated resonance (see below). For CO$_2$,  comparing HITRAN
intensities with our predictions we would expect the same, but
detailed analysis (cf. figure \ref{figure:Int_20012}), that such jumps
in accuracy cause artificial patterns in line intensities
within a single vibrational band. 

All HITRAN2012 entries taken from a pre-release version of CDSD have
been tagged with uncertainty code 3 (20\%\ or worse). 
However, this number does not reflect actual uncertainties of
the intensities. Most of the HITRAN intensities have the uncertainties
much better than 20\%. More detailed information about the actual
uncertainties can be found in the official release of CDSD
\cite{15TaPeGa.CO2}. The reader should use this work in order to get a
realistic information about the uncertainties of the line parameters.

\begin{figure}[H]
\includegraphics[width=12cm]{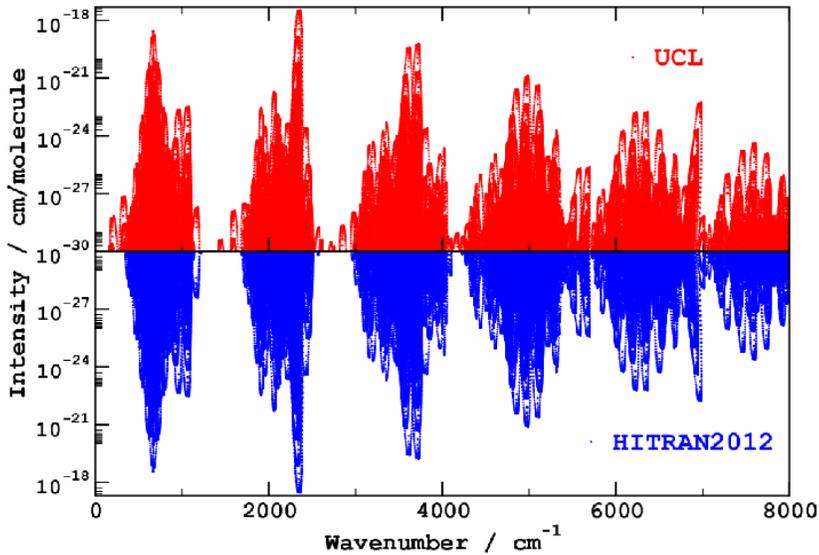}
\caption{General comparison of the HITRAN2012 and UCL CO$_2$ 296~K linelists for the 0 -- 8000 \cm\ region.}
\centering
\label{figure:overview}
\end{figure}

Figure \ref{figure:overview} gives a general overview of the two
linelists. Overall the agreement is excellent with more than 98\%\ of
entries common between both lists and very similar intensities. However,
there is some incomplete coverage by HITRAN2012
with several artificial windows, especially for low intensity
transitions. There are also a few missing medium-intensity
transitions around 400 \cm\ and 1600 \cm. 

%\subsection{Intensities. General overview.}

\begin{figure}[H]
\includegraphics[width=12cm]{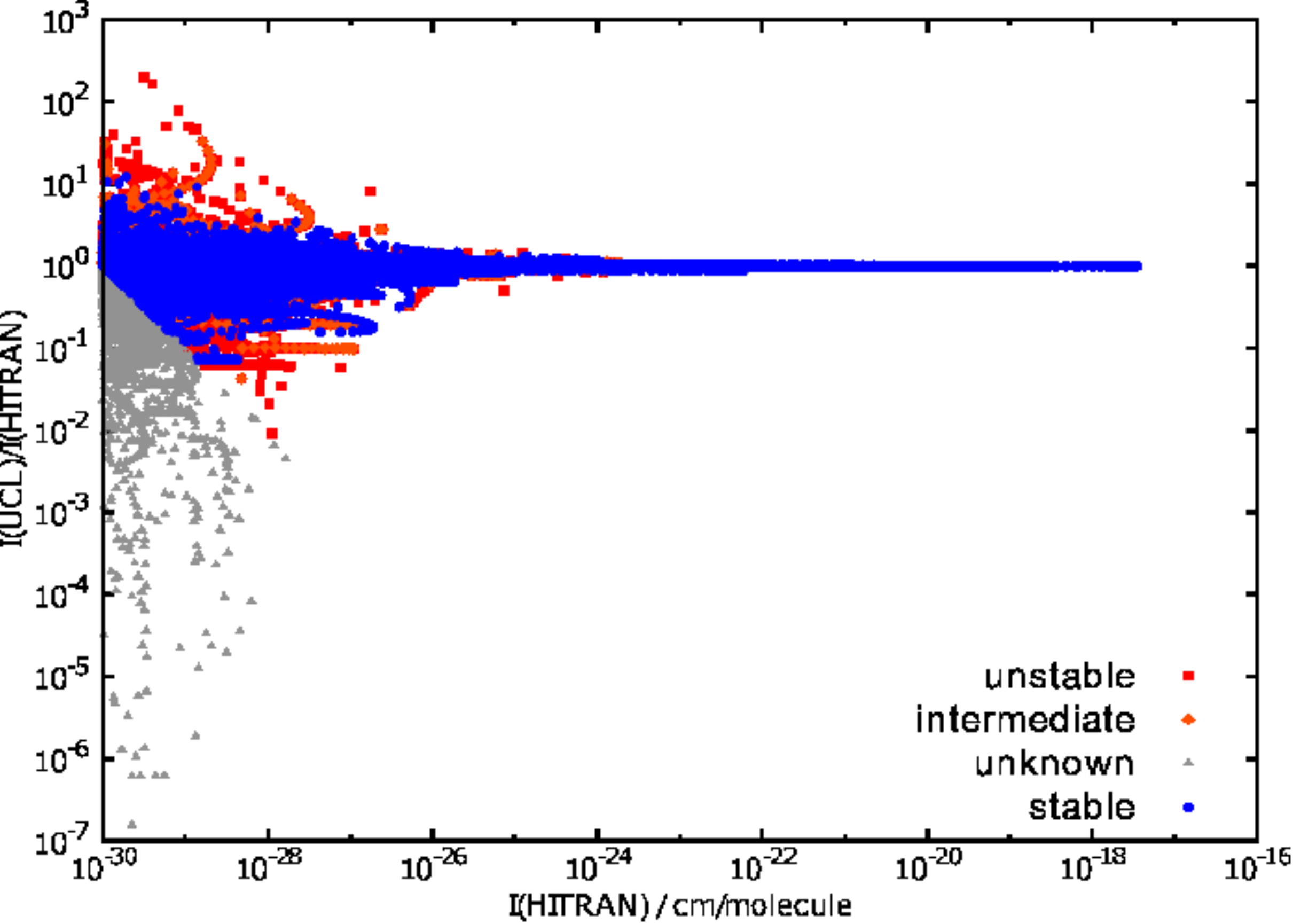}
\caption{Comparison of HITRAN2012 and UCL line intensities: UCL to
  HITRAN intensity ratio as a function of HITRAN line intensity. Blue
  points stand for unstable lines according to our sensitivity
  analysis, while red points are considered to be stable.  There are
  147 000 stable, 7000 intermediate, 4400 unstable and 1 400 unknown
  lines which are too weak for a scatter factor to be determined reliably.}
\centering
\label{figure:fullcomp}
\end{figure}

Intensities of all assigned UCL lines relative to HITRAN2012 are
depicted in figure ~\ref{figure:fullcomp}. As expected discrepancies between the two
linelists grow as lines get weaker, which results in a funnel-like shape
in the plot which characteristic of such comparisons (e.g. \cite{jt285}). 
The stability of the UCL lines on the scatter factors are
also shown; as could be  anticipated stable lines predominate 
at higher intensities.

%\subsection{Intensities. HITRAN uncertainty regions.}

It is instructive to divide HITRAN2012 data into subsets of a given
intensity accuracy code. Each of those sets can be then compared to
our results separately to provide an estimate for compatibility of two
linelists at different levels of accuracy. To achieve that we plotted 
HITRAN intensities with the  accuracy code found for CO$_2$ which is 7 
(2~\%\ or better uncertainty) against the UCL ones.
This set of lines encompass the important
20011, 20012, 20013, 30011, 30012, 30013 and 30014 bands as well as  the asymmetric
stretching second overtone 00031. All bands except intermediate 30011 band are stable.  Comparisons with high accuracy measurements above have
already shown that our results for the 30013 band are accurate to about
1~\%\ or better.

%  Similarly we obtain sub-percent agreement with
% recent high quality experiment for the 00031 band; both our calculations
% and these new measurements seem to deviate from the
% measurements of Toth \etal\ \cite{08ToBrMi} by 10\%; this
% is also consistent with our difference from CSDD-296 for this band.

Again one can see
characteristic bow-like structures corresponding to particular
rotational transitions within a vibrational band, with the peak of an
arc refers to most intense, low $J$ transition. We suggest that these
structures are artifacts which originate from the semi-empirical treatment of the intensities.

\begin{figure}[H]
\centering
\includegraphics[width=12cm]{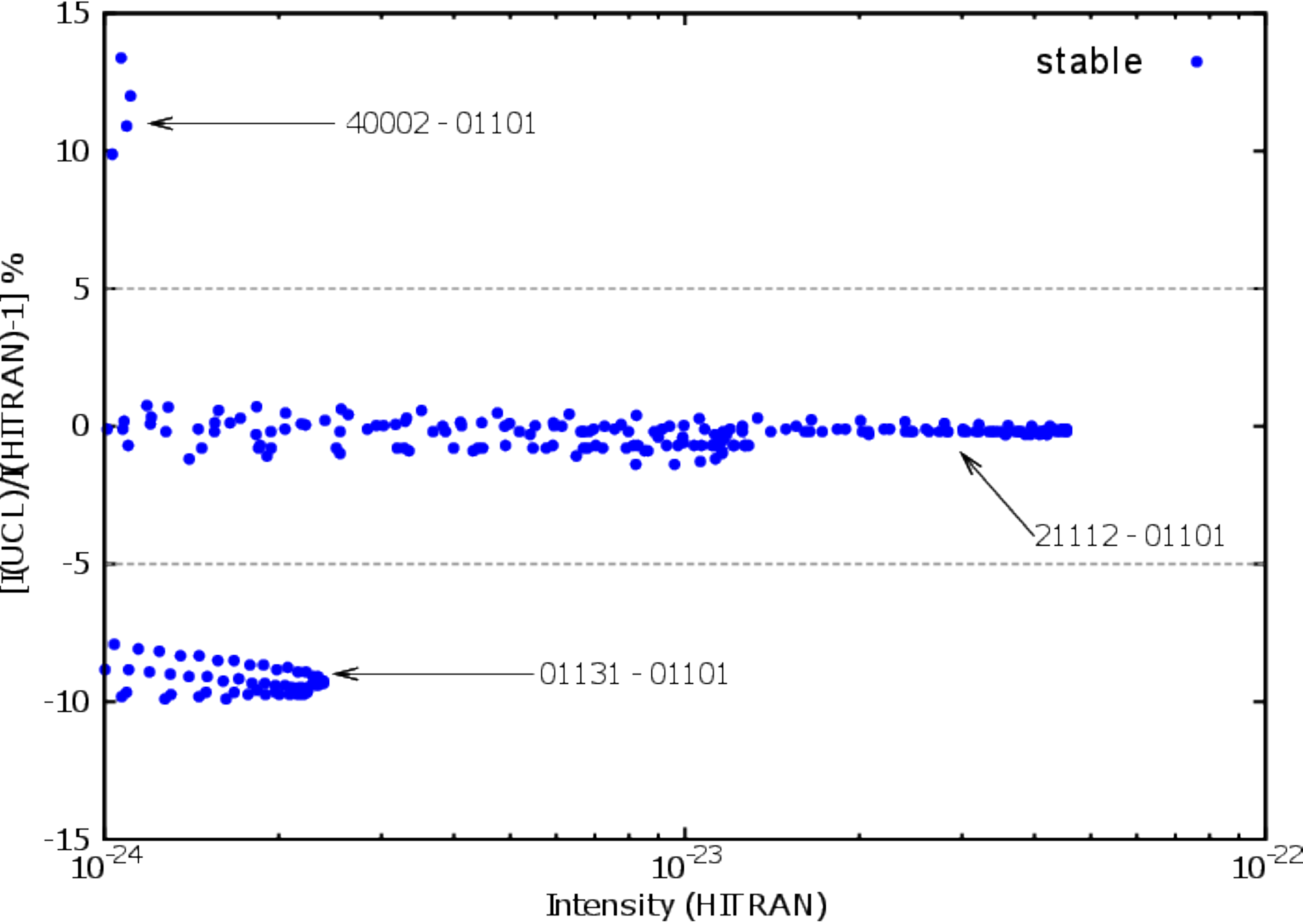}
\caption{Comparison of HITRAN2012 medium-accuracy intensities and UCL line intensities. 
The dashed line indicates the stated HITRAN uncertainty, i.e. 5\%. Arrows label vibrational
bands.}
\label{figure:ier6}
\end{figure}

A similar situation occurs for bands with HITRAN uncertainty code 6, see 
figure \ref{figure:ier6}; here very good agreement is spoiled by
01131 -- 01101 band.

\begin{figure}[H]
\centering
\includegraphics[width=12cm]{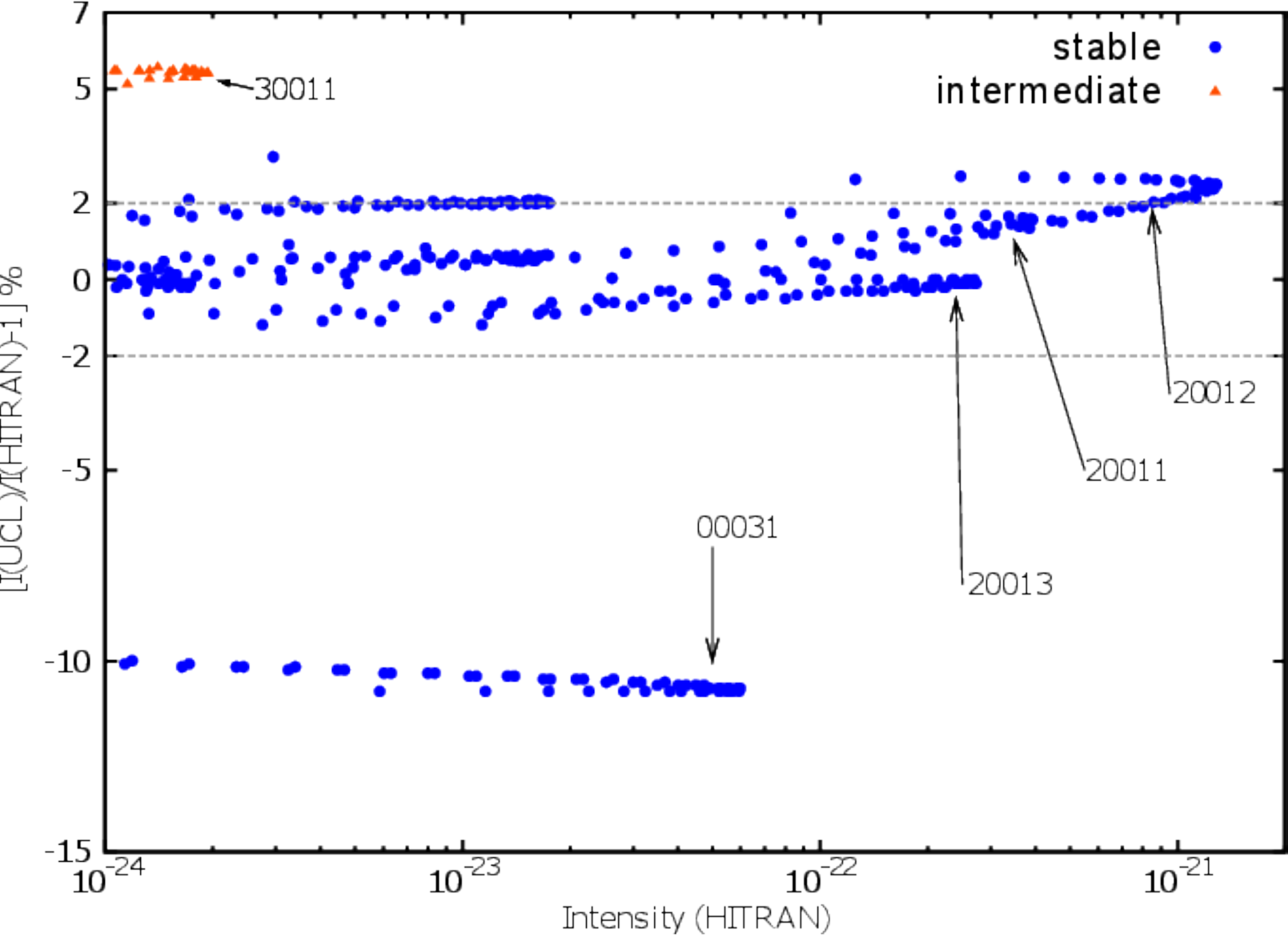}
\caption{Comparison of HITRAN2012 most accurate intensities and UCL line intensities. 
The dashed line indicates the stated HITRAN uncertainty, i.e. 2\%. Arrows label vibrational
bands, which all start from the ground 00001 state.}
\label{figure:toth}
\end{figure}

%\subsection{Intensities. Selected bands.}

Figure~\ref{figure:Int_20012} gives an intensities comparison for the
20012 band.  HITRAN 2012 used two separate data sources for this band.
This is clearly visible which means, despite the overall good
agreement with present results, there is an abrupt change in intensity
trends at $J=64$. This is the point where the experimental data
finished and the database had to rely on results from a the CDSD
effective Hamiltonian calculations.

\subsection{A HITRAN-style line list}
The final UCL-296 line list, given as supplementary data, contains
162~468 line positions, intensities scaled by natural abundance
(0.98420), quantum numbers and scatter factors taken from our
computation.

Our final recommended line list for $^{12}$C$^{16}$O$_2$ is
also given in supplementary data. This list contains
162~260 lines in HITRAN format
with intensities scaled by natural abundance and uniformly cut off at
10$^{-30}$ cm/molecule. Vibrational quantum numbers were taken from
CDSD  and cross-checked with HITRAN2012 assignments. 
Line positions were transferred from CDSD-296 with appropriate uncertainties.  

The majority of line intensities (151~602) were taken from our ``AU``
calculations; we assign HITRAN uncertainty code 8 for stable bands
with at least one transition stronger than $10^{-23}$ cm/molecule and
7 for stable bands weaker than $10^{-25}$ cm/molecule together with
8647 lines from intermediate bands.
 
Whenever our line intensity turned out to be unreliable (i.e. was either
unstable and no additional tests confirmed its high accuracy or belonged to 
$3v_3$ family of bands) it was replaced by CDSD-296 value.
This was the case for  10 080 (6\%) lines.

\section{Conclusion}

We present a new mixed ab initio-empirical linelist providing reliable
intensities for $^{12}$C$^{16}$O$_2$  up to 8000 \cm. 
We believe  this line list is more complete and the intensities more
accurate than in HITRAN2012 \cite{jt557}. 
A detailed analysis
shows that our line intensities generally are accurate
at the  sub-percent level when compared
to recent, high-accuracy measurements, consequently validating
our approach; furthermore we find that intensity  uncertainties 
stated in HITRAN2012 are probably too conservative. 
We believe these improved intensities
should assist to improve CO$_2$ monitoring in
 remote atmospheric sensing studies, and in other applications.
Furthermore this new line lists fills in the small gaps in the
 HITRAN2012 list. Of course for use in atmospheric conditions
this line list needs to be supplemented by both line profile
parameters and consideration of line-mixing \cite{10LaTrLa.CO2}.
   
One issue that we should raise concerns perpendicular transitions
(those with $\Delta \ell = \pm 1$ and $\pm 2$).The majority of the perpendicular
bands borrow intensity from the considerably stronger parallel
($\Delta \ell=0$) bands via Coriolis resonance or anharmonic plus
$\ell$-type interactions.  To describe this process it is necessary to
have very precise wavefunctions.  So far, the very high accuracy of
the line intensity calculations presented here is confirmed
experimentally only for parallel bands. All weaker bands have been
given a lower accuracy rating in our line list; none-the-less it would
be very helpful to have some high accuracy experimental measurements
of perpendicular bands to help to independently validate our results.

Future work will focus on two aspects of the problem. First, it is
apparent that our {\it ab initio} dipole moment surface is less
accurate for transitions involving changes of 3 or more quanta in $\nu_3$.
This problem will be the subject of future theoretical investigation which
will also aim to extend our model to frequencies higher than 8000 \cm.
Second, a major advantage of our methodology is that theoretical calculations
can be used to give intensities for all isotopologues of CO$_2$ with essentially
the same accuracy as the  $^{16}$O$^{12}$C$^{16}$O results presented
here. This should be particularly used in providing accurate intensities
for trace species such as  $^{16}$O$^{14}$C$^{16}$O, which are important
for monitoring purposes \cite{13LeMiWo.CO2}. Line lists for isotopically
substituted CO$_2$ will be published in the near future.

\section*{Acknowledgments}
This work is supported by the UK Natural Environment Research Council (NERC) through grant NE/J010316,
the ERC under the Advanced Investigator Project
267219 and the Russian Fund for Fundamental Science.  
We would like to thank Jens Brunzendorf and V Malathy Devi for sharing their results with us prior to publication.

\bibliographystyle{elsarticle-num}

%\bibliography{journals_phys,jtj,CO2,waternewsources,methods}

\begin{thebibliography}{10}
\expandafter\ifx\csname url\endcsname\relax
  \def\url#1{\texttt{#1}}\fi
\expandafter\ifx\csname urlprefix\endcsname\relax\def\urlprefix{URL }\fi
\expandafter\ifx\csname href\endcsname\relax
  \def\href#1#2{#2} \def\path#1{#1}\fi

\bibitem{GOSAT}
A.~Butz, S.~Guerlet, O.~Hasekamp, D.~Schepers, A.~Galli, I.~Aben,
  C.~Frankenberg, J.~M. Hartmann, H.~Tran, A.~Kuze, G.~Keppel-Aleks, G.~Toon,
  D.~Wunch, P.~Wennberg, N.~Deutscher, D.~Griffith, R.~Macatangay,
  J.~Messerschmidt, J.~Notholt, T.~Warneke, {Toward accurate CO$_2$ and CH$_4$
  observations from GOSAT}, Geophys. Res. Lett. {38} ({2011}) L14812.
\newblock \href {http://dx.doi.org/{10.1029/2011GL047888}}
  {\path{doi:{10.1029/2011GL047888}}}.

\bibitem{OCO}
D.~Crisp, R.~Atlas, F.~Breon, L.~Brown, J.~Burrows, P.~Ciais, B.~Connor,
  S.~Doney, I.~Y. Fung, D.~Jacob, C.~Miller, D.~O'Brien, S.~Pawson, J.~T.
  Randerson, P.~Rayner, R.~Salawitch, S.~Sander, B.~Sen, G.~Stephens, P.~Tans,
  G.~Toon, P.~Wennberg, S.~Wofsy, Y.~Yung, Z.~Kuang, B.~Chudasama, G.~Sprague,
  B.~Weiss, R.~Pollock, D.~Kenyon, S.~Schroll, {The Orbiting Carbon Observatory
  (OCO) mission}, Adv. Space Res. {34} ({2004}) 700--709.
\newblock \href {http://dx.doi.org/{10.1016/j.asr.2003.08.062}}
  {\path{doi:{10.1016/j.asr.2003.08.062}}}.

\bibitem{ASCENDS}
J.~B. Abshire, H.~Riris, G.~R. Allan, C.~J. Weaver, J.~Mao, X.~Sun, W.~E.
  Hasselbrack, A.~Yu, A.~Amediek, Y.~Choi, E.~V. Browell, {A Lidar Approach to
  Measure CO$_2$ Concentrations from Space for the ASCENDS Mission}, {SPIE}
  {7832} ({2010}) 78320D.
\newblock \href {http://dx.doi.org/{10.1117/12.868567}}
  {\path{doi:{10.1117/12.868567}}}.

\bibitem{TCCON}
D.~Wunch, G.~C. Toon, J.-F.~L. Blavier, R.~A. Washenfelder, J.~Notholt, B.~J.
  Connor, D.~W.~T. Griffith, V.~Sherlock, P.~O. Wennberg, {The Total Carbon
  Column Observing Network}, Phil. Trans. Royal Soc. London A {369} ({2011})
  2087--2112.
\newblock \href {http://dx.doi.org/{10.1098/rsta.2010.0240}}
  {\path{doi:{10.1098/rsta.2010.0240}}}.

\bibitem{NDACC}
F.~Hase, {Improved instrumental line shape monitoring for the ground-based,
  high-resolution FTIR spectrometers of the Network for the Detection of
  Atmospheric Composition Change}, Atmos. Meas. Tech. {5} ({2012}) 603--610.

\bibitem{Emmert2012}
J.~T. Emmert, M.~H. Stevens, P.~F. Bernath, D.~P. Drob, C.~D. Boone,
  Observations of increasing carbon dioxide concentration in earth's
  thermosphere, Nature Geoscience 5 (2012) 868--871.
\newblock \href {http://dx.doi.org/10.1038/ngeo1626}
  {\path{doi:10.1038/ngeo1626}}.

\bibitem{jt453}
L.~S. Rothman, I.~E. Gordon, A.~Barbe, D.~C. Benner, P.~F. Bernath, M.~Birk,
  V.~Boudon, L.~R. Brown, A.~Campargue, J.~P. Champion, K.~Chance, L.~H.
  Coudert, V.~Dana, V.~M. Devi, S.~Fally, J.~M. Flaud, R.~R. Gamache,
  A.~Goldman, D.~Jacquemart, I.~Kleiner, N.~Lacome, W.~J. Lafferty, J.~Y.
  Mandin, S.~T. Massie, S.~N. Mikhailenko, C.~E. Miller, N.~Moazzen-Ahmadi,
  O.~V. Naumenko, A.~V. Nikitin, J.~Orphal, V.~I. Perevalov, A.~Perrin,
  A.~Predoi-Cross, C.~P. Rinsland, M.~Rotger, M.~Simeckova, M.~A.~H. Smith,
  K.~Sung, S.~A. Tashkun, J.~Tennyson, R.~A. Toth, A.~C. Vandaele,
  J.~Vander~Auwera, {The {\it HITRAN} 2008 molecular spectroscopic database},
  J. Quant. Spectrosc. Radiat. Transf. 110 (2009) 533--572.

\bibitem{jt584}
J.~Tennyson, P.~F. Bernath, A.~Campargue, A.~G. Cs\'asz\'ar, L.~Daumont, R.~R.
  Gamache, J.~T. Hodges, D.~Lisak, O.~V. Naumenko, L.~S. Rothman, H.~Tran,
  N.~F. Zobov, J.~Buldyreva, C.~D. Boone, M.~D. {De Vizia}, L.~Gianfrani, J.-M.
  Hartmann, R.~McPheat, J.~Murray, N.~H. Ngo, O.~L. Polyansky, D.~Weidmann,
  {Recommended isolated-line profile for representing high-resolution
  spectroscopic transitions (IUPAC Technical Report)}, Pure Appl. Chem. 86
  (2014) 1931--1943.

\bibitem{Wang2005}
L.~Wang, V.~I. Perevalov, S.~A. Tashkun, Y.~Ding, S.-M. Hu, {Absolute line
  intensities of $^{13}$C$^{16}$O$_2$ in the 4200{\textendash}8500 cm$^{-1}$
  region}, J. Mol. Spectrosc. 234 (2005) 84--92.
\newblock \href {http://dx.doi.org/10.1016/j.jms.2005.08.008}
  {\path{doi:10.1016/j.jms.2005.08.008}}.

\bibitem{Perevalov2008}
B.~V. Perevalov, A.~Campargue, B.~Gao, S.~Kassi, S.~A. Tashkun, V.~I.
  Perevalov, New {CW}-{CRDS} measurements and global modeling of
  {$^{12}$C$^{16}$O$_2$} absolute line intensities in the 1.6$\mu$m region, J.
  Mol. Spectrosc. 252 (2008) 190--197.
\newblock \href {http://dx.doi.org/10.1016/j.jms.2008.08.006}
  {\path{doi:10.1016/j.jms.2008.08.006}}.

\bibitem{Song2010}
K.~F. Song, S.~Kassi, S.~A. Tashkun, V.~I. Perevalov, A.~Campargue, High
  sensitivity {CW}-cavity ring down spectroscopy of 12co2 near 1.35$\mu$m
  ({II}): New observations and line intensities modeling, J. Quant. Spectrosc.
  Radiat. Transf. 111 (2010) 332--344.
\newblock \href {http://dx.doi.org/10.1016/j.jqsrt.2009.09.004}
  {\path{doi:10.1016/j.jqsrt.2009.09.004}}.

\bibitem{Hashemi2013}
R.~Hashemi, H.~Rozario, A.~Ibrahim, A.~Predoi-Cross, {Line shape study of the
  carbon dioxide laser band I$^1$}, Can. J. Phys. 91 (2013) 924--936.
\newblock \href {http://dx.doi.org/10.1139/cjp-2013-0051}
  {\path{doi:10.1139/cjp-2013-0051}}.

\bibitem{06BoMaDa.CO2}
D.~Boudjaadar, J.-Y. Mandin, V.~Dana, N.~Picqu{\'e}, G.~Guelachvili,
  {$^{12}$C$^{16}$O$_2$ line intensity measurements around 1.6 $\mu$m}, J. Mol.
  Spectrosc. 236 (2006) 158 -- 167.
\newblock \href {http://dx.doi.org/http://dx.doi.org/10.1016/j.jms.2006.01.007}
  {\path{doi:http://dx.doi.org/10.1016/j.jms.2006.01.007}}.

\bibitem{OBrien2002}
D.~M. O'Brien, P.~J. Rayner, {Global observations of the carbon budget, 2,
  CO$_2$ column from differential absorption of reflected sunlight in the 1.61
  $\mu$m band of CO$_2$}, J. Geophys. Res. 107~(D18) (2002) ACH~6.
\newblock \href {http://dx.doi.org/10.1029/2001jd000617}
  {\path{doi:10.1029/2001jd000617}}.

\bibitem{XOCO}
C.~E. Miller, D.~Crisp, P.~L. DeCola, S.~C. Olsen, J.~T. Randerson, A.~M.
  Michalak, A.~Alkhaled, P.~Rayner, D.~J. Jacob, P.~Suntharalingam, D.~B.~A.
  Jones, A.~S. Denning, M.~E. Nicholls, S.~C. Doney, S.~Pawson, H.~Boesch,
  B.~J. Connor, I.~Y. Fung, D.~O'Brien, R.~J. Salawitch, S.~P. Sander, B.~Sen,
  P.~Tans, G.~C. Toon, P.~O. Wennberg, S.~C. Wofsy, Y.~L. Yung, R.~M. Law,
  {Precision requirements for space-based X-CO2 data}, J. Geophys. Res. {112}
  ({2007}) D10314.
\newblock \href {http://dx.doi.org/{10.1029/2006JD007659}}
  {\path{doi:{10.1029/2006JD007659}}}.

\bibitem{Sioris2014}
C.~E. Sioris, C.~D. Boone, R.~Nassar, K.~J. Sutton, I.~E. Gordon, K.~A. Walker,
  P.~F. Bernath, Retrieval of carbon dioxide vertical profiles from solar
  occultation observations and associated error budgets for {ACE}-{FTS} and
  {CASS}-{FTS}, Atmos. Meas. Tech. 7 (2014) 2243--2262.
\newblock \href {http://dx.doi.org/10.5194/amt-7-2243-2014}
  {\path{doi:10.5194/amt-7-2243-2014}}.

\bibitem{07CaPaCa.CO2}
G.~Casa, D.~A. Parretta, A.~Castrillo, R.~Wehr, L.~Gianfrani, Highly accurate
  determinations of co$_2$ line strengths using intensity-stabilized diode
  laser absorption spectrometry, J. Chem. Phys. 127 (2007) 084311.
\newblock \href {http://dx.doi.org/http://dx.doi.org/10.1063/1.2759930}
  {\path{doi:http://dx.doi.org/10.1063/1.2759930}}.

\bibitem{09CaWeCa.CO2}
G.~Casa, R.~Wehr, A.~Castrillo, E.~Fasci, L.~Gianfrani, {The line shape problem
  in the near-infrared spectrum of self-colliding CO$_2$ molecules:
  Experimental investigation and test of semiclassical models}, J. Chem. Phys.
  130 (2009) 184306.
\newblock \href {http://dx.doi.org/http://dx.doi.org/10.1063/1.3125965}
  {\path{doi:http://dx.doi.org/10.1063/1.3125965}}.

\bibitem{11WuViJo.CO2}
G.~Wuebbeler, G.~J.~P. Viquez, K.~Jousten, O.~Werhahn, C.~Elster, {Comparison
  and assessment of procedures for calculating the R(12) line strength of the
  $\nu_1 +2\nu_2 + \nu_3$ band of CO$_2$}, J. Chem. Phys. {135} ({2011})
  204304.
\newblock \href {http://dx.doi.org/{10.1063/1.3662134}}
  {\path{doi:{10.1063/1.3662134}}}.

\bibitem{jt613}
O.~L. Polyansky, K.~Bielska, M.~Ghysels, L.~Lodi, N.~F. Zobov, J.~T. Hodges,
  J.~Tennyson, {High accuracy CO$_2$ line intensities determined from theory
  and experiment}, Phys. Rev. Lett. 114 (2015) 243001.
\newblock \href {http://dx.doi.org/10.1103/PhysRevLett.114.24300}
  {\path{doi:10.1103/PhysRevLett.114.24300}}.

\bibitem{Malathy2015}
V.~M. Devi, D.~C. Benner, K.~Sung, L.~R. Brown, T.~J. Crawford, C.~E. Miller,
  B.~J. Drouin, V.~H. Payne, S.~Yu, M.~A.~H. Smith, A.~W. Mantz, Line
  parameters including temperature dependences of self- and air-broadened line
  shapes of {$^{12}$C$^{16}$O$_2$}: 1.6-$\mu$m region, J. Quant. Spectrosc.
  Radiat. Transf.\href {http://dx.doi.org/submitted} {\path{doi:submitted}}.

\bibitem{15BrWeSe.CO2}
J.~Brunzendorf, V.~Werwein, A.~Serduykov, O.~Werhahn, V.~Ebert, {CO$_2$ line
  strength measurements in the 20012--00001 band near 2 $\mu$m}, in: The 24th
  Colloquium on High Resolution Molecular Spectroscopy, 2015, p. O17.

\bibitem{Tashkun2015}
S.~A. Tashkun, V.~I. Perevalov, R.~R. Gamache, J.~Lamouroux, {CDSD}-296, high
  resolution carbon dioxide spectroscopic databank: Version for atmospheric
  applications, J. Quant. Spectrosc. Radiat. Transf. 152 (2015) 45--73.
\newblock \href {http://dx.doi.org/10.1016/j.jqsrt.2014.10.017}
  {\path{doi:10.1016/j.jqsrt.2014.10.017}}.

\bibitem{92WaRoxx.CO2}
R.~B. Wattson, L.~S. Rothman, Direct numerical diagonalization - wave of the
  future, J. Quant. Spectrosc. Radiat. Transf. 48 (1992) 763--780.

\bibitem{94DaMaBa.CO2}
V.~Dana, J.~Y. Mandin, A.~Barbe, J.~J. Plateaux, L.~S. Rothman, R.~B. Wattson,
  {$^{12}$C$^{16}$O$_2$} line-intensities in the 4.8 $\mu$m spectral region, J.
  Quant. Spectrosc. Radiat. Transf. {52} ({1994}) 333--340.
\newblock \href {http://dx.doi.org/10.1016/0022-4073(94)90163-5}
  {\path{doi:10.1016/0022-4073(94)90163-5}}.

\bibitem{12HuScTa.CO2}
X.~Huang, D.~W. Schwenke, S.~A. Tashkun, T.~J. Lee, {An isotopic-independent
  highly accurate potential energy surface for CO$_2$ isotopologues and an
  initial $^{12}$C$^{16}$O$_2$ infrared line list}, J. Chem. Phys. 136 (2012)
  124311.
\newblock \href {http://dx.doi.org/10.1063/1.3697540}
  {\path{doi:10.1063/1.3697540}}.

\bibitem{13HuFrTa.CO2}
X.~Huang, R.~S. Freedman, S.~A. Tashkun, D.~W. Schwenke, T.~J. Lee,
  {Semi-empirical $^{12}$C$^{16}$O$_2$ IR line lists for simulations up to 1500
  K and 20,000 cm$^{-1}$}, J. Quant. Spectrosc. Radiat. Transf. {130} ({2013})
  134--146.
\newblock \href {http://dx.doi.org/10.1016/j.jqsrt.2013.05.018}
  {\path{doi:10.1016/j.jqsrt.2013.05.018}}.

\bibitem{14HuGaFr.CO2}
X.~Huang, R.~R. Gamache, R.~S. Freedman, D.~W. Schwenke, T.~J. Lee, {Reliable
  infrared line lists for 13 CO$_2$ isotopologues up to E =18,000 cm$^{-1}$ and
  1500 K, with line shape parameters}, J. Quant. Spectrosc. Radiat. Transf.
  {147} ({2014}) 134--144.
\newblock \href {http://dx.doi.org/{10.1016/j.jqsrt.2014.05.015}}
  {\path{doi:{10.1016/j.jqsrt.2014.05.015}}}.

\bibitem{92TeSuPe.CO2}
J.~L. Teffo, O.~N. Sulakshina, V.~I. Perevalov, Effective {H}amiltonian for
  rovibrational energies and line-intensities of carbon-dioxide, J. Mol.
  Spectrosc. {156} ({1992}) 48--64.
\newblock \href {http://dx.doi.org/10.1016/0022-2852(92)90092-3}
  {\path{doi:10.1016/0022-2852(92)90092-3}}.

\bibitem{02TaPeTe.CO2}
S.~A. Tashkun, V.~I. Perevalov, J.~L. Teffo, A.~D. Bykov, N.~N. Lavrentieva,
  {CDSD-1000, the high-temperature carbon dioxide spectroscopic databank}, J.
  Quant. Spectrosc. Radiat. Transf. {82} ({2003}) 165--196.
\newblock \href {http://dx.doi.org/{10.1016/S0022-4073(03)00152-3}}
  {\path{doi:{10.1016/S0022-4073(03)00152-3}}}.

\bibitem{11TaPe.CO2}
S.~A. Tashkun, V.~I. Perevalov, {CDSD-4000: High-resolution, high-temperature
  carbon dioxide spectroscopic databank}, J. Quant. Spectrosc. Radiat. Transf.
  112 (2011) 1403--1410.
\newblock \href {http://dx.doi.org/10.1016/j.jqsrt.2011.03.005}
  {\path{doi:10.1016/j.jqsrt.2011.03.005}}.

\bibitem{jt509}
L.~Lodi, J.~Tennyson, O.~L. Polyansky, A global, high accuracy ab initio dipole
  moment surface for the electronic ground state of the water molecule, J.
  Chem. Phys. 135 (2011) 034113.

\bibitem{jt522}
L.~Lodi, J.~Tennyson, {Line lists for H$_2$$^{18}$O and H$_2$$^{17}$O based on
  empirically-adjusted line positions and ab initio intensities}, J. Quant.
  Spectrosc. Radiat. Transf. 113 (2012) 850--858.

\bibitem{jt530}
M.~Grechko, O.~Aseev, T.~R. Rizzo, N.~F. Zobov, L.~Lodi, J.~Tennyson, O.~L.
  Polyansky, O.~V. Boyarkin, {Stark coefficients for highly excited
  rovibrational states of H$_2$O}, J. Chem. Phys. 136 (2012) 244308.

\bibitem{jt587}
A.~Petrignani, M.~Berg, A.~Wolf, I.~I. Mizus, O.~L. Polyansky, J.~Tennyson,
  N.~F. Zobov, M.~Pavanello, L.~Adamowicz, Visible intensities of the triatomic
  hydrogen ion from experiment and theory, J. Chem. Phys. 141 (2014) 241104.
\newblock \href {http://dx.doi.org/10.1063/1.4904440}
  {\path{doi:10.1063/1.4904440}}.

\bibitem{jt557}
L.~S. Rothman, I.~E. Gordon, Y.~Babikov, A.~Barbe, D.~C. Benner, P.~F. Bernath,
  M.~Birk, L.~Bizzocchi, V.~Boudon, L.~R. Brown, A.~Campargue, K.~Chance, E.~A.
  Cohen, L.~H. Coudert, V.~M. Devi, B.~J. Drouin, A.~Fayt, J.-M. Flaud, R.~R.
  Gamache, J.~J. Harrison, J.-M. Hartmann, C.~Hill, J.~T. Hodges,
  D.~Jacquemart, A.~Jolly, J.~Lamouroux, R.~J. {Le Roy}, G.~Li, D.~A. Long,
  O.~M. Lyulin, C.~J. Mackie, S.~T. Massie, S.~Mikhailenko, H.~S.~P.
  M{\"u}ller, O.~V. Naumenko, A.~V. Nikitin, J.~Orphal, V.~Perevalov,
  A.~Perrin, E.~R. Polovtseva, C.~Richard, M.~A.~H. Smith, E.~Starikova,
  K.~Sung, S.~Tashkun, J.~Tennyson, G.~C. Toon, V.~G. Tyuterev, G.~Wagner, {The
  {\it HITRAN} 2012 molecular spectroscopic database}, J. Quant. Spectrosc.
  Radiat. Transf. 130 (2013) 4 -- 50.
\newblock \href {http://dx.doi.org/10.1016/jqsrt.2013.07.002}
  {\path{doi:10.1016/jqsrt.2013.07.002}}.

\bibitem{14ReOuMiWa}
L.~Regalia, C.~Oudot, S.~Mikhailenko, L.~Wang, X.~Thomas, A.~Jenouvrier,
  P.~Von~der Heyden, Water vapor line parameters from 6450 to 9400 cm$^{-1}$,
  J. Quant. Spectrosc. Radiat. Transf. {136} ({2014}) 119--136.
\newblock \href {http://dx.doi.org/{10.1016/j.jqsrt.2013.11.019}}
  {\path{doi:{10.1016/j.jqsrt.2013.11.019}}}.

\bibitem{08ToBrMi}
R.~A., L.~R. Brown, C.~E. Miller, V.~M. Devi, D.~C. Benner, Spectroscopic
  database of $\mathrm{CO_2}$ line parameters: 4300-7000 $\mathrm{cm^{-1}}$, J.
  Quant. Spectrosc. Radiat. Transf. 109 (2008) 906--921.

\bibitem{12TaPe}
S.~A. Tashkun, V.~I. Perevalov (2012).

\bibitem{jt45}
B.~T. Sutcliffe, J.~Tennyson, A generalised approach to the calculation of
  ro-vibrational spectra of triatomic molecules, Mol. Phys. 58 (1986)
  1053--1066.

\bibitem{jt96}
B.~T. Sutcliffe, J.~Tennyson, A general treatment of vibration-rotation
  coordinates for triatomic molecules, Intern. J. Quantum Chem. 39 (1991)
  183--196.

\bibitem{jt114}
J.~Tennyson, B.~T. Sutcliffe, Discretisation to avoid singularities in
  vibration-rotation hamiltonians: a bisector embedding for {AB$_2$}
  triatomics, Intern. J. Quantum Chem. 42 (1992) 941--952.

\bibitem{jt160}
J.~Tennyson, J.~R. Henderson, N.~G. Fulton, {DVR3D}: programs for fully
  pointwise calculation of ro-vibrational spectra of triatomic molecules,
  Comput. Phys. Commun. 86 (1995) 175--198.

\bibitem{jt338}
J.~Tennyson, M.~A. Kostin, P.~Barletta, G.~J. Harris, O.~L. Polyansky,
  J.~Ramanlal, N.~F. Zobov, {DVR3D: a program suite for the calculation of
  rotation-vibration spectra of triatomic molecules}, Comput. Phys. Commun. 163
  (2004) 85--116.

\bibitem{jt156}
A.~E. {Lynas-Gray}, S.~Miller, J.~Tennyson, Infra red transition intensities
  for water: a comparison of {\it ab initio} and fitted dipole moment surfaces,
  J. Mol. Spectrosc. 169 (1995) 458--467.

\bibitem{12WeKnKn.methods}
H.-J. Werner, P.~J. Knowles, G.~Knizia, F.~R. Manby, M.~Sch\"utz, Molpro: a
  general-purpose quantum chemistry program package, WIREs Comput. Mol. Sci. 2
  (2012) 242--253.
\newblock \href {http://dx.doi.org/10.1002/wcms.82}
  {\path{doi:10.1002/wcms.82}}.

\bibitem{jt475}
L.~Lodi, J.~Tennyson, {Theoretical methods for small-molecule ro-vibrational
  spectroscopy}, J. Phys. B: At. Mol. Opt. Phys. 43 (2010) 133001.

\bibitem{jt14}
J.~Tennyson, B.~T. Sutcliffe, {The ab initio calculation of the
  vibration-rotation spectrum of triatomic systems in the close-coupling
  approach with KCN and H$_2$Ne as examples}, J. Chem. Phys. 77 (1982)
  4061--4072.

\bibitem{jtDown}
M.~J. Down, L.~Lodi, J.~Tennyson, {Line lists for HD$^{16}$O, HD$^{18}$O and
  HD$^{17}$O based on empirical line positions and {\it ab initio}
  intensities}, J. Quant. Spectrosc. Radiat. Transf.

\bibitem{jt454}
J.~Tennyson, P.~F. Bernath, L.~R. Brown, A.~Campargue, M.~R. Carleer, A.~G.
  Cs\'asz\'ar, R.~R. Gamache, J.~T. Hodges, A.~Jenouvrier, O.~V. Naumenko,
  O.~L. Polyansky, L.~S. Rothman, R.~A. Toth, A.~C. Vandaele, N.~F. Zobov,
  L.~Daumont, A.~Z. Fazliev, T.~Furtenbacher, I.~E. Gordon, S.~N. Mikhailenko,
  S.~V. Shirin, {IUPAC critical Evaluation of the Rotational-Vibrational
  Spectra of Water Vapor. Part I. Energy Levels and Transition Wavenumbers for
  H$_2$$^{17}$O and H$_2$$^{18}$O}, J. Quant. Spectrosc. Radiat. Transf. 110
  (2009) 573--596.

\bibitem{jt482}
J.~Tennyson, P.~F. Bernath, L.~R. Brown, A.~Campargue, M.~R. Carleer, A.~G.
  Cs\'asz\'ar, L.~Daumont, R.~R. Gamache, J.~T. Hodges, O.~V. Naumenko, O.~L.
  Polyansky, L.~S. Rothman, R.~A. Toth, A.~C. Vandaele, N.~F. Zobov, A.~Z.
  Fazliev, T.~Furtenbacher, I.~E. Gordon, S.~N. Mikhailenko, B.~A. Voronin,
  {IUPAC critical Evaluation of the Rotational-Vibrational Spectra of Water
  Vapor. Part II. Energy Levels and Transition Wavenumbers for HD$^{16}$O,
  HD$^{17}$O, and HD$^{18}$O}, J. Quant. Spectrosc. Radiat. Transf. 111 (2010)
  2160--2184.

\bibitem{jt562}
J.~Tennyson, P.~F. Bernath, L.~R. Brown, A.~Campargue, A.~G. Cs\'asz\'ar,
  L.~Daumont, R.~R. Gamache, J.~T. Hodges, O.~V. Naumenko, O.~L. Polyansky,
  L.~S. Rothman, A.~C. Vandaele, N.~F. Zobov, {A Database of Water Transitions
  from Experiment and Theory (IUPAC Technical Report)}, Pure Appl. Chem. 86
  (2014) 71--83.

\bibitem{jt412}
T.~Furtenbacher, A.~G. {Cs\'asz\'ar}, J.~Tennyson, {MARVEL: measured active
  rotational-vibrational energy levels}, J. Mol. Spectrosc. 245 (2007)
  115--125.

\bibitem{12FuCsi.method}
T.~Furtenbacher, A.~G. {Cs\'asz\'ar}, {MARVEL: measured active
  rotational-vibrational energy levels. II. Algorithmic improvements}, J.
  Quant. Spectrosc. Radiat. Transf. 113 (2012) 929--935.

\bibitem{04AmViCh.CO2}
A.~Amy-Klein, H.~Vigu{\'e}, C.~Chardonnet, {Absolute frequency measurement of
  $^{12}$C$^{16}$O$_2$ laser lines with a femtosecond laser comb and new
  determination of the $^{12}$C$^{16}$O$_2$ molecular constants and frequency
  grid}, J. Mol. Spectrosc. 228 (2004) 206--212.

\bibitem{13PoOtWe.CO2}
A.~Pogany, O.~Ott, O.~Werhahn, V.~Ebert, {Towards traceability in CO$_2$ line
  strength measurements by TDLAS at 2.7 $\mu m$}, J. Quant. Spectrosc. Radiat.
  Transf. {130} ({2013}) 147--157.
\newblock \href {http://dx.doi.org/{10.1016/j.jqsrt.2013.07.011}}
  {\path{doi:{10.1016/j.jqsrt.2013.07.011}}}.

\bibitem{jt309}
O.~L. Polyansky, A.~G. {Cs\'asz\'ar}, S.~V. Shirin, N.~F. Zobov, P.~Barletta,
  J.~Tennyson, D.~W. Schwenke, P.~J. Knowles, {High accuracy ab initio
  rotation-vibration transitions of water}, Science 299 (2003) 539--542.

\bibitem{15TaPeGa.CO2}
S.~A. Tashkun, V.~I. Perevalov, R.~R. Gamache, J.~Lamouroux, {CDSD}-296, high
  resolution carbon dioxide spectroscopic databank: Version for atmospheric
  applications, J. Quant. Spectrosc. Radiat. Transf. 152 (2015) 45--73.
\newblock \href {http://dx.doi.org/10.1016/j.jqsrt.2014.10.017}
  {\path{doi:10.1016/j.jqsrt.2014.10.017}}.

\bibitem{jt285}
R.~Schermaul, R.~C.~M. Learner, A.~A.~D. Canas, J.~W. Brault, O.~L. Polyansky,
  D.~Belmiloud, N.~F. Zobov, J.~Tennyson, Weak line water vapor spectrum in the
  regions 13~200 -- 15~000 cm$^{-1}$, J. Mol. Spectrosc. 211 (2002) 169--178.

\bibitem{10LaTrLa.CO2}
J.~Lamouroux, H.~Tran, A.~L. Laraia, R.~R. Gamache, L.~S. Rothman, I.~E.
  Gordon, J.~M. Hartmann, {Updated database plus software for line-mixing in
  CO$_2$2 infrared spectra and their test using laboratory spectra in the
  1.5-2.3 $\mu$m region}, J. Quant. Spectrosc. Radiat. Transf. {111} ({2010})
  2321--2331.
\newblock \href {http://dx.doi.org/{10.1016/j.jqsrt.2010.03.006}}
  {\path{doi:{10.1016/j.jqsrt.2010.03.006}}}.

\bibitem{13LeMiWo.CO2}
S.~J. Lehman, J.~B. Miller, C.~Wolak, J.~Southon, P.~P. Tans, S.~A. Montzka,
  C.~Sweeney, A.~Andrews, B.~LaFranchi, T.~P. Guilderson, J.~C. Turnbull,
  Allocation of terrestrial carbon sources using {$^{14}$CO$_2$}: Methods,
  measurement, and modeling, Radiocarbon {55} ({2013}) 1484--1495.

\end{thebibliography}

\end{document}